\documentclass[acmsmall,manuscript]{acmart}
\usepackage{multirow}
\usepackage{array}
\usepackage{graphicx}
\AtBeginDocument{%
  \providecommand\BibTeX{{%
    \normalfont B\kern-0.5em{\scshape i\kern-0.25em b}\kern-0.8em\TeX}}}

\setcopyright{acmlicensed}
\acmJournal{PACMHCI}
\acmYear{2025} \acmVolume{xx} \acmNumber{CSCW1} \acmArticle{xx} \acmMonth{xx} \acmPrice{15.00}\acmDOI{10.1145/xxxxxxx}

\definecolor{activegold}{RGB}{255,193,61}
\definecolor{lightorange}{RGB}{230, 170, 50}
\definecolor{lightgreen}{RGB}{121,210,121}
\definecolor{lightteal}{RGB}{121,199,210}
\definecolor{lightblue}{RGB}{100,212,239}
\definecolor{lightpurple}{RGB}{153,102,255}

\usepackage{xspace}
\usepackage{xcolor}

\newcommand{\pzh}[1]{{\color{black} #1}}
\newcommand{\peng}[1]{{\color{black} #1}}
\newcommand{\zhenhui}[1]{{\color{black} #1}}

\newcommand{\tw}[1]{{\color{black} #1}}
\newcommand{\taewook}[1]{{\color{black} #1}}
\newcommand{\rl}[1]{{\color{black} #1}}

\newcommand{\update}[1]{{\color{black} #1}}
\newcommand{\rv}[1]{{\color{black} #1}}

\newcommand{\name}{{\textit{MentalImager}}}

\newcommand{\ie}{\textit{i.e.,} }
\newcommand{\eg}{\textit{e.g.,} }
\newcommand{\etal}{\textit{et al.} }

\begin{document}

\title{MentalImager: Exploring Generative Images for Assisting Support-Seekers' Self-Disclosure in Online Mental Health Communities}

\author{Han Zhang}
\email{zhangh773@mail2.sysu.edu.cn}
\affiliation{%
  \institution{Sun Yat-sen University}
  \city{Zhuhai}
  \country{China}
}

\author{Jiaqi Zhang}
\email{zhangjq93@mail2.sysu.edu.cnn}
\affiliation{%
  \institution{Sun Yat-sen University}
  \city{Zhuhai}
  \country{China}
}

\author{Yuxiang Zhou}
\email{zhouyx95@mail2.sysu.edu.cn}
\affiliation{%
  \institution{Sun Yat-sen University}
  \city{Zhuhai}
  \country{China}
}

\author{Ryan Louie}
\email{rylouie@cs.stanford.edu}
\affiliation{%
  \institution{Stanford University}
  \city{Stanford}
  \country{United States}
}

\author{Taewook Kim}
\email{taewook@u.northwestern.edu}
\affiliation{%
  \institution{Northwestern University}
  \city{Evanston}
  \country{United States}
}

\author{Qingyu Guo}
\email{qguoag@connect.ust.hk}
\affiliation{%
  \institution{Hong Kong University of Science and Technology}
  \city{Hong Kong}
  \country{China}
}

\author{Shuailin Li}
\email{lishlin8@mail2.sysu.edu.cn}
\affiliation{%
  \institution{Sun Yat-sen University}
  \city{Zhuhai}
  \country{China}
}

\author{Zhenhui Peng}
\authornote{Corresponding author.}
\email{pengzhh29@mail.sysu.edu.cn}
\affiliation{%
  \institution{Sun Yat-sen University}
  \city{Zhuhai}
  \country{China}
}

\renewcommand{\shortauthors}{Han Zhang, et al.}

\begin{abstract}
Support-seekers’ self-disclosure of their suffering experiences, thoughts, and feelings in the post can help them get needed peer support in online mental health communities (OMHCs). However, such mental health self-disclosure could be challenging. Images can facilitate the manifestation of relevant experiences and feelings in the text; yet, relevant images are not always available. In this paper, we present a technical prototype named MentalImager \zhenhui{and validate in a human evaluation study that it} can generate topical- and emotional-relevant images based on the seekers' drafted posts or specified keywords. \zhenhui{
Two user studies demonstrate that MentalImager not only improves seekers' satisfaction with their self-disclosure in their posts but also invokes support-providers' empathy for the seekers and willingness to offer help. 
Such improvements are credited to the generated images, which help seekers express their emotions and inspire them to add more details about their experiences and feelings.
}
We report concerns on MentalImager and discuss insights for supporting self-disclosure in OMHCs.
\end{abstract}






\maketitle

\section{Introduction}
\tw{People experiencing emotional difficulties (e.g., anxiety, sadness, grief, depression) often turn to online to seek social support from peers.}
Online Mental Health Communities (OMHCs) \tw{(e.g., Reddit's r/depression and r/GriefSupport) can} offer an easily accessible and anonymous way that assists individuals in \tw{receiving such} support from peers \cite{Pavalanathan15, Pennebaker96}. 
\tw{For example, users in OMHCs can share their personal experiences and emotions through posts in text and optionally with images~\cite{Peng21, Li16, Moorhead13, Wang23, Ding23}. 
Such self-disclosure of feelings, thoughts, and experiences is key to the realization of OMHCs' benefits for satisfying support-seekers' needs for others' social support and improving their well-being \cite{Peng21, Li16, Luo20}.
}
\tw{
On the one hand, proper self-disclosure can provide insightful information about the communicator \cite{Haker05}, support impression management practices \cite{Wang16}, and facilitate the development of social relationships \cite{Tamir12}. 
On the other hand, posts with poorly articulated or unclear self-disclosure tend to result in receiving less effective informational and emotional support from peers within these communities~\cite{Yang19}.}
\rl{While receiving social support on OMHCs depends on the quality of self-disclosure, not all individuals find it easy to describe their experiences, thoughts, and feelings in their posts~\cite{Leary17}.}
Researchers in Human-Computer Interaction(HCI) have explored various methods like \rl{an} AI-assisted writing tool \cite{Shin22} and chatbot \cite{Lee20} to guide users \rl{in writing their thoughts and experiences} on OMHCs. 
These methods help mental health self-disclosure in pure text but overlook other modalities (e.g., images, videos, audio) that can be used in self-disclosure. 

For example, it is a common practice for people to attach images in their posts related to mental health on social media \cite{Feuston19, Zhao21, Manikonda17}, as images help express complicated emotions and thoughts about mental health issues \cite{Leary17, Kaplan10, Manikonda17}.
\rl{However, in cases of no proper images in local devices, it can be time-consuming to draw a picture or search for an image that can reveal a support-seeker’s thoughts and feelings. A potential solution for using images in self-disclosure posts lies in text-to-image generative AI models, which can quickly and accurately transform written descriptions into visual representations. }
These techniques have been applied in art creation \cite{Wang2023} and game development \cite{Tilson19} but have not been explored for facilitating online mental health self-disclosure. 
\rl{Current HCI literature lacks insight into user needs for generated images in mental health self-disclosure, the effectiveness of existing generative techniques for this task, and their impact on self-disclosure and eliciting support in OMHCs.}
Such understandings are important as they can provide design considerations for researchers to develop generative techniques for healthcare scenarios, inform OMHCs to provide adaptive assistance to mediate their members in communication and help support-seekers improve their well-being. 

\begin{figure}[H]
  \centering
  \includegraphics[width=1\textwidth]{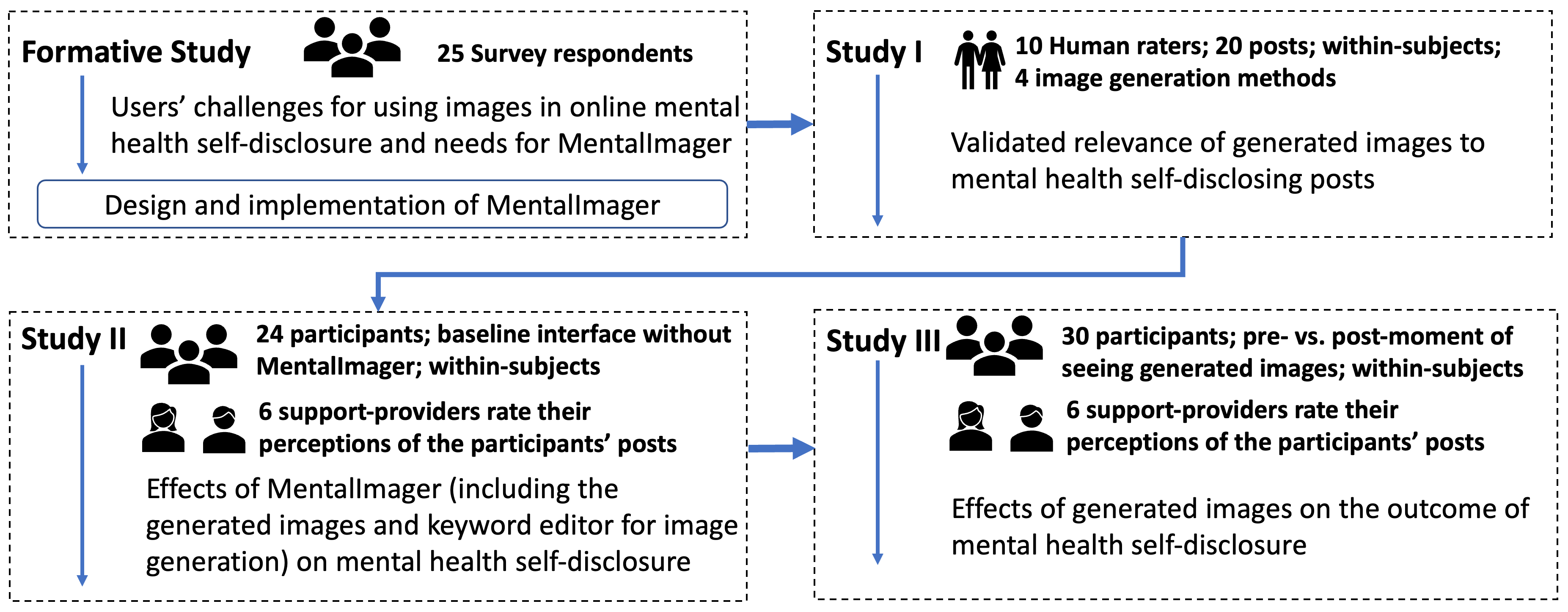} 
  \caption{\zhenhui{Our studies to design and evaluate MentalImager for facilitating mental health self-disclosure with generated images in online communities.}}
  \label{fig:study_flow}
\end{figure}

\peng{
In this paper, we design, \zhenhui{develop, and evaluate} a technical prototype named MentalImager that uses text-to-image generation techniques to facilitate mental health self-disclosure in online communities. 
\zhenhui{\autoref{fig:study_flow} summarizes our design and evaluation process of MentalImager.} 
We first \rl{distributed a survey to 25 respondents to understand their practices and challenges when using images to disclose their mental health issues online, and preferences if using text-to-image generative models for this task.}
Respondents highlighted the benefits of images for helping them disclose their issues, difficulties in finding proper images, and expectations for a tool that can interactively generate topical- and emotional-relevant images given drafted posts or keywords. 
Based on these findings, we presented MentalImager as an assistant for drafting support-seeking posts on a simulated OMHC web page. 
Users can write down initial thoughts in text, click a button to generate related images based on the text, manipulate the detected keywords of the text and their weights to regenerate the images, re-edit the text, attach a generated image to the post, and finally submit the post to the simulated OMHC. 
}

We conducted \zhenhui{one human evaluation study} to explore RQ1) whether and how text-to-image generative techniques can generate images that are topical- and emotional-relevant to support-seeking posts in OMHCs, \zhenhui{as well as two user studies to evaluate RQ2)} how would \zhenhui{MentalImager} impact the outcome and experience of self-disclosure via the posts in OMHCs. 
\zhenhui{First, to address RQ1, we conducted a Study I with 10 human raters judging the relevance between the text of support-seeking posts in OMHCs and the corresponding images generated by} 
a state-of-the-art Stable Diffusion text-to-image generative model used in MentalImager. 
Specifically, we proposed to use the extracted keywords and detected emotion of the post content as the prompt to the image generative model. 
\zhenhui{Our Study I found that the generated images using such a prompt were rated with higher topical- and emotional-relevance to the text, compared to the images generated using two traditional textual prompts, \ie full text of the post and extracted keywords of the post, and images retrieved from the Google Image Search engine. 
This result validates the feasibility of generating relevant images to the support-seeking posts in OMHCs. 
}

\zhenhui{
Then, we conducted two within-subjects studies, \ie Study II and Study III, to address RQ2.
Study II (N = 24) compared MentalImager with a baseline interface that provides neither generated images nor editable keywords. 
The results indicate that MentalImager significantly improves users' satisfaction with their support-seeking posts and decreases their difficulties with mental health self-disclosure. 
Participants commonly re-edited the textual content of their posts after viewing the generated images and editing the keywords. }
We also invited another six people to act as support-providers to rate their perception of the posts from the user study. 
We found that providers were more empathetic towards and more willing to respond to posts assisted by MentalImager. 
\zhenhui{
Nevertheless, Study II could not tell the effects of generated images from the editable keywords in MentalImager as participants appreciated both features. 
Therefore, Study III (N = 30) specifically compared the self-disclosure outcome at pre-moment, \ie after checking the editable keywords but before seeing the generated images for the first time, with the self-disclosure outcome at post-moment, \ie after seeing the generated images and submitting the post. 
The results show that participants deemed their mental health self-disclosure significantly more clear at post-moment than at pre-moment, and they had significantly less worry about being misunderstood by providers at post-moment. 
Providers also had a significantly higher degree of empathy and understanding of the posts at post-moment compared to those at pre-moment. 
Participants indicated that the generated images helped them express their emotions, represented the content of the text in their posts, and could attract viewers' attention to their posts.  
}
We also report participants' concerns on MentalImager and conclude with design considerations for using generated images to facilitate self-disclosure in online mental health communities.

Our work makes three contributions to the CSCW community. 
First, we present a technical prototype named MentalImager to facilitate online mental health self-disclosure via generated images. 
Second, our \zhenhui{two user studies} demonstrates that MentalImager helps to improve support-seekers' satisfaction and self-disclosure in their posts and gives first-hand findings on how they interact and perceive the generated images in this process. 
Third, we showcase the feasibility of generative models for generating images that are topical- and emotional-relevant to the mental health self-disclosure text and offer design considerations for leveraging generative AIs to assist support-seekers in OMHCs.

\section{Related work}
\subsection{Self-Disclosure in Online Mental Health Communities}
    Online mental health communities (OMHCs) offer a convenient way for individuals to access social support \cite{Newman11, Chancellor19} and has been demonstrated to be helpful for alleviating their distress and fulfilling psychological needs \cite{Choudhury17, Gleeson12}. 
    The exchange of social support is a primary motivator for users to join OMHCs \cite{Choudhury14, Pavalanathan2015}. 
    In OMHCs, support-seekers can anonymously express their emotions or share sensitive experiences \cite{Choudhury14, Pavalanathan2015}, which is considered as mental health self-disclosure \cite{Martin11}.
    
    
    Numerous studies (e.g., \cite{Yang19, Lith16, Wang16}) have substantiated the advantageous effects of self-disclosure on improving people's mental health. Through self-disclosure, individuals can alleviate stress and engage in self-examination, which could serve as a form of self-therapy \cite{Ma21}. 
    Proper self-disclosure can facilitate the development of social relationship with others and help support-seekers get social support \cite{Tamir12}. 
  For example, \citet{Yang19} analyzed the posts and comments in Cancer Survivor Network and found that one's self-disclosure leads others to self-disclose and to provide support. 
   Specifically, support-seekers' negative self-disclosure was associated with the perception that they were seeking both emotional and informational support which, in turn, was associated with them receiving the corresponding type of support \cite{Yang19}. 
    However, disclosing the experience and feelings about the mental health issues could be challenging for distressed individuals. 
    For instance, while accurate or authentic disclosure may correspond to positive well-being \cite{grieve2016psychological}, distressed people are more likely to present the ``false self'' on social media \cite{gil2015facebook} and thus may feel even more depressed and less connected.
    The support-seekers' mental health status \cite{Kim08} and their lack of experience and understanding of the OMHCs \cite{Song23} may also impair their self-disclosure process and outcome. 
    
    
    
    Our research is motivated by the benefits of self-disclosure for improving support-seekers' mental health in OMHCs and aims to maximize these benefits by augmenting the self-disclosure process through AI-generated images. 
    


    \subsection{Facilitating Communication in Online Mental Health Communities}

To benefit members in OMHCs, online platforms have explored various approaches to facilitate positive communication among peers. 
    One approach is to provide assistance manually, such as training members with therapeutic skills to assist support-seekers \cite{Rodgers10}. 
    However, this approach is often time-consuming and difficult to be applied at a larger scale. 
    Another approach is to use AI techniques to improve the community environment, such as building classifiers to automatically filter out negative content from the community for users \cite{Dosono13} and predict inappropriate posts \cite{Song23}. 
    Nevertheless, this method is more for moderators of the communities and does not offer direct assistance to the members during their communication. 

Our work is more aligned with the approach of AI-mediated communication \rl{between support providers and help seekers.}
From the support-providers' perspectives, \citet{Peng20} modeled the supportiveness of comments in an OMHC and leverage these models to power MepsBot that assesses the users' drafted comments and recommends relevant supportive comments for reference. 
They show that MepsBot can improve support-providers' satisfaction with their comments and improve the supportiveness of the comments to the support-seeking posts \cite{Peng20}. 
In the scenario of online peer-to-peer counseling platform, \citet{diyi_modeling_cscw22} modeled the motivational interviewing strategies of support-providers, which can power an AI tool that helps providers build understanding with seekers \cite{supporter_challenges_cscw21}. 
From the support-seekers' perspectives, prior researchers have explored how to guide users disclose their thoughts \rl{and} feelings. 
For example, \citet{Lee20} investigated how chatbots can be used to prompt deep self-disclosure and found that users were more willing to share sensitive information with chatbots than with mental health professionals, thereby increasing their desire for sharing their information. 
\citet{Shin22} designed a system that displays contextual keywords to help support-seekers clarify emotion and describe text concretely while writing a post. 
Such assistants can guide \rl{users in  formulating their thoughts during} the writing process \cite{Jakesch23}. 

However, words may be insufficient for expressing certain experiences or emotions \cite{Manikonda17}.
As an option, many OMHCs allow support-seekers to attach images in their posts, which may help them convey emotions and thoughts about mental health issues \cite{Leary17, Kaplan10, Manikonda17,Zhao14}. 
For example, \citet{Manikonda17} extracted the visual features (e.g., color), themes, and emotions relating to mental health disclosures on Instagram, indicating the use of imagery for unique self-disclosure need. 
Nevertheless, little work explores how to facilitate seekers to use images in their self-disclosure and how the images may affect support-providers' perceptions on the posts. 
Our work seeks to bridge these gaps by exploring the potentials of generative images for facilitating mental health self-disclosure. 
Our results provide implications of using generative images to mediate peer support communication.

    \subsection{Image Generation Techniques}

Image generation techniques have become a popular topic in the fields of AI and HCI. 
These techniques can produce realistic images based on user input in various domains, such as art and medical imaging.
    As for art, generative models like DeepDream and StyleGAN have been used to create artistic and surreal images \cite{She20}. 
    As for medical imaging, generative adversarial networks (GANs) can be used to enhance medical imaging, such as generating high-resolution MRI or CT scan images from lower-resolution inputs. 
    This can aid in diagnosis and treatment planning \cite{Batiste04}. 
    Currently, mainstream models fall into two categories: those that take images as input and those that take text as input. 
    CycleGAN \cite{Gu20}, for example, is a GAN-based model that typically takes images as input and can learn relationships between images in different domains, enabling image transformations for tasks like cross-domain image translation and artistic creation. 
    DALL-E \cite{Yu22}, on the other hand, is a Transformer-based model that can generate images based on textual descriptions, commonly used for tasks involving automated image generation.
    Our work chooses to employ a text-to-image model as our image generation method as the support-seeking posts in online mental health communities are normally text-based. 
    Specifically, we selected the open-source Stable Diffusion model that can produce different images based on various types of textual inputs (\eg keywords, sentences) \cite{She20, Gu20, Yu22, Batiste04}. 
    
    Previous work on the text-to-image generation models normally examine the models' performance on generating images that are relevant to the content or topic of the input text. 
    However, it often overlooks whether the image and text are relevant regarding their emotions, which could be helpful for disclosing mentally challenging issues \cite{Belkofer14, Martin11}. 
    Besides, it is under-investigated how the generated images would impact support-seekers' mental health self-disclosure process and outcome in OMHCs. 
    In this paper, we provide answers to these two questions via Studies I and  II. 

\section{Design and Implementation of MentalImager}
\subsection{Challenges in Mental Health Self-Disclosure and Needs for MentalImager}
\label{sec:survey}
\peng{
To understand users' challenges for using images in online mental health self-disclosure and needs for MentalImager, we distributed a survey and collected results from 25 respondents. 
}

\peng{
\subsubsection{Survey Protocol and Respondents}
\update{We developed our questionnaire on WenJuanXing \footnote{WenJuanXing (\url{https://www.wjx.cn/}) is a common platform for creating questionnaires in China.}. 
We chose to distribute the questionnaire to university students, who often face mental health challenges and seek social support online \cite{mental_health_college_chi20,mental_health_university_student} and would be the representative users of MentalImager. 
The inclusion criteria is that respondents should have experiences in mental health self-disclosure online. 
}
Therefore, in the questionnaire, we first asked a yes-or-no question to filter out respondents who have no experience in disclosing thoughts and feelings of mental health challenges online. 
\tw{
Then we asked our respondents to describe their recent experiences in seeking social support via open-ended text submissions. We particularly asked about their challenges in self-disclosure through text and/or images.  Also, we \rl{asked them} what kind of topics are challenging to express through such modalities (i.e., text and images) for seeking social support online.}
Next, we asked a) the level of agreement (1 - strongly disagree, 7 - strongly agree) on the helpfulness of adding images in mental health self-disclosure and b) the difficulties when they want to use images in the self-disclosure via a multiple-choice question with the options brainstormed by our research team, i.e., ``no proper image in the local device'', ``can not find proper images online'', ``can find proper images online but it is time-consuming'', ``no idea about what images are proper'', others. 
Finally, we used example-generated images to introduce the text-to-image generative techniques in brief. 
We asked respondents to rate the perceived usefulness (1 - least useful, 7 - very useful) of an assistant powered by these generative techniques for helping them in self-disclosure and its three potential features brainstormed by our research team for providing generated images, i.e., based on the post's textual content, the emotion revealed in the post, and specified keywords. 

We recruited 25 students (S1-25, 7 \textbf{M}ales, 11 \textbf{F}emales, 7 \textbf{N}ot to specify) through social networks in the local university. 
Their ages range from 19 to 24 years old with an average age of 20.72 (SD = 1.64). 
Eleven respondents described their tough times about school work, e.g., \textit{``My advisor talked to me yesterday. I was afraid that I would be expelled from school. I was depressed and had a headache''} (S12, F, age: 21).
Eight respondents talked about their relationship with their friends or families, e.g., ``I used to disclose my experience in Weibo about how my mom made me depressed and led to my stomach disorders. People nearby normally have a good family and can not empathize with me. There are more people on the Internet who have the same family problems as me'' (S21, N, age: 20). 
Others mainly shared their frustrating experiences with the application of jobs (N = 4) and sleep (3). 
}

\peng{
    \subsubsection{Findings}
\label{subsub:findings}
We summarized the topics that respondents found difficult to disclose their negative experience and emotions in pure text and tend to add images: 
romantic relationship (N = 8),
current situation and emotion (7), 
school work (6), 
and 
family (3).
Respondents generally agreed that it is helpful to add images to their posts for disclosing mental health experiences and emotions (Mean = 4.96, SD = 1.95). 
Among the challenges in the multiple-choices question, 9 respondents chose ``no proper image in the local device'', 5 ``can not find proper images online'', 15 ``can find proper images online but it is time-consuming'', and 10 ``no idea about what images are proper''. 
This result indicates that it is time-consuming and often unavailable for users to find an appropriate existing image for their posts about mental health self-disclosure. 
As for the perceived usefulness of the potential features, 19 out of the 25 respondents agree that it would be useful (score $\geq$ 5 in the 7-point Likert scale) to have an assistant that can generate images to help them in self-disclosure.
In general, respondents rate it useful if the assistant could generate images based on the post's textual content (17 / 25),
the emotion revealed in the post (17 / 25),
and user-specified keywords (18 / 25). 
These results suggest the user needs for MentalImager to interactively generate topical- and emotional-relevant images based on the user's drafted post or specified keywords. 



}

\peng{
\subsection{Design of MentalImager}
\tw{Given the findings in~\autoref{subsub:findings}, we designed our MentalImager as illustrated in~\autoref{fig:interface} a--e.} 
Following~\cite{Peng20}, we chose to embed MentalImager as a sidebar into the interface of existing online mental health communities (OMHCs) so that the target users could easily access its support on the community website.
As a demonstration, we simulate an anonymous OMHC for people who are depressed or struggling with depression. 
Note that we position MentalImager as a non-commercial tool to facilitate self-disclosure of mental health concerns in online communities. 
Users do not necessarily need to attach the generated images to their posts nor re-edit their posts if they do not feel like doing so.
    
\begin{figure}[h]
  \centering
  \includegraphics[width=\textwidth]{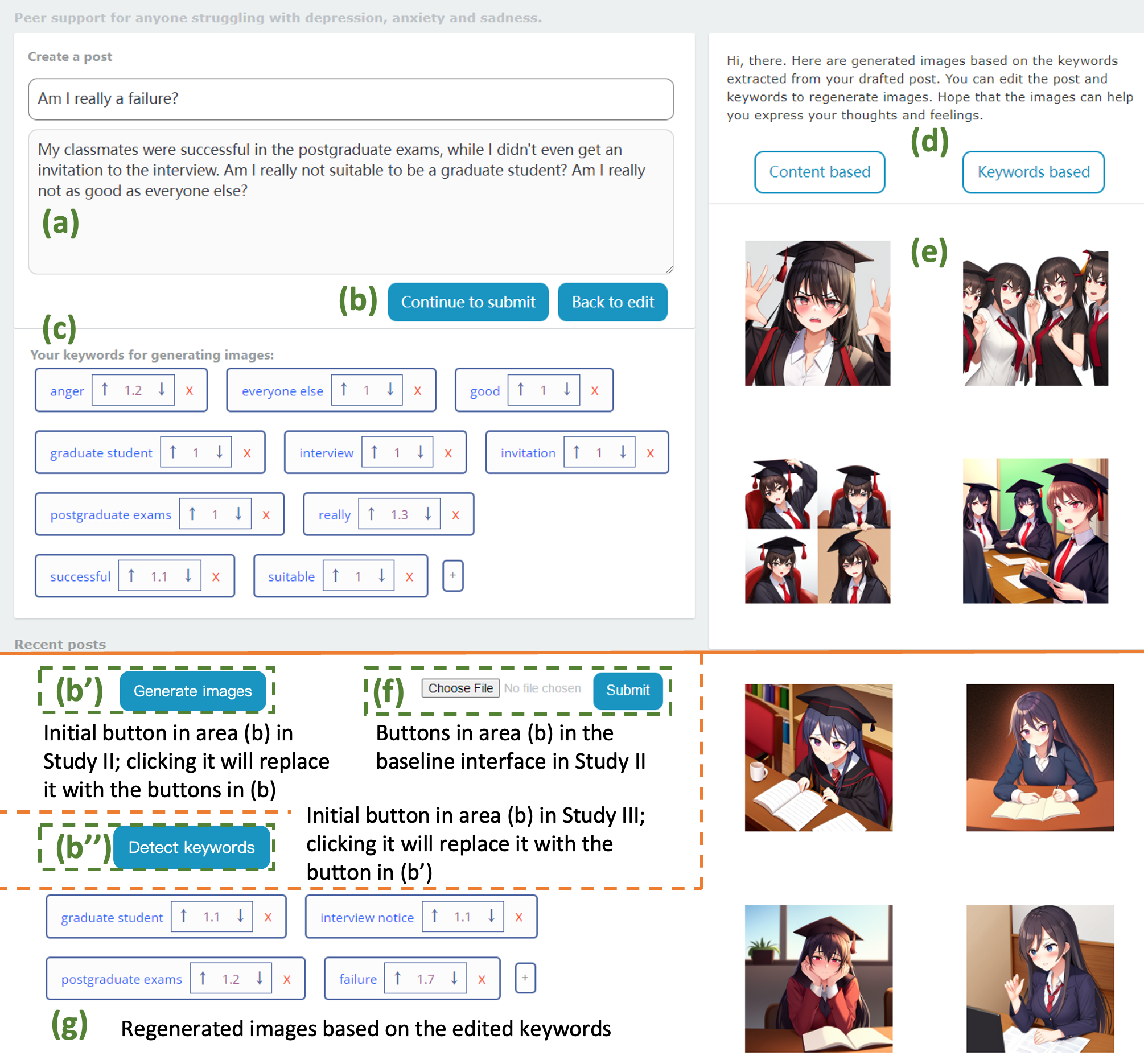} 
  \caption{
  Screenshot (parts a-e) of the MentalImager embedded in a simulated community. 
  (a) Area for creating a post. \zhenhui{(b, b', and b'') Buttons for detecting keywords (only appear in Study III), generating images, continuing to submit, and re-editing the post.} (c) Keyword editor panel. (d and e) Panel for regenerating and displaying images. (f) Buttons for uploading an image and submitting the post in Study II's baseline interface. (g) Regenerated images based on the edited keywords. 
  }
  \label{fig:interface}
\end{figure}

\subsubsection{Interaction with MentalImager}
\label{sec:user_interface}


To create a support-seeking post with MentalImager, users can write the very first draft in the text areas (\autoref{fig:interface}a). 
They can click the ``Generate images'' button (b') to invoke MentalImager to generate images relevant to the current draft. 
The interface will then show up the \update{keyword editor panel} (c) and the image displaying panel (d,e), and the ``Generate images'' button will be replaced by the ``Continue to submit'' and ``Back to edit'' buttons (b). 
Such a ``Generate images <-> (Back to edit) -> Continue to submit'' submission flow is inspired by \citet{Peng20} and could mitigate interruption in users' composing process. 
Users can choose ``Back to edit'' to refine the text in their posts or ``Continue to submit'' to formally publish the content whenever they want. \rv{In this case, users will be unable to edit the text content until they click ``Back to edit'' button, which is designed to keep users inputting text while extracting keywords to avoid read-write conflicts. }

In the keyword editor panel (\autoref{fig:interface}c), 
users can check the keywords extracted from their posts and the weights of these keywords used for the text-to-image generation model. 
Users can click each keyword to edit its content, click ``$\uparrow$''/``$\downarrow$'' to increase/decrease the weight by 0.1 \rv{(default increasing trend of model)} or directly edit the number of the weight, and click ``x''/``+'' to delete/add a keyword. \rv{The extraction and initial weights of the keywords are introduced in the following \autoref{sec:prompts}.}
Inspired by \citet{Shin22}, such interactive keyword tags could help users iteratively generate needed images and refine their drafted posts. 
\autoref{fig:interface}g shows an example of users' editions of the keywords and the regenerated images based on the edited keywords.
\zhenhui{The flexible keyword editor allow users to modify, delete, and add keywords and adjust their weights for image generation, which means that it allows users to input any keywords, e.g., about age, gender, occupation, major, etc. to guide the model to generate related images. 
The keyword detection module can also extract keywords about the poster's own characteristics if they appear in the self-disclosure posts. For example, in the example shown in \autoref{fig:interface}c, MentalImager detects the keyword ``graduate student'' from the drafted post and generates images related to the graduate student (e).}

In the image displaying panel (\autoref{fig:interface}d,e), users can check four generated images, which could increase the possibility that users like one of the images while avoiding overwhelming users. 
By default, the images are generated using the extracted keywords (c) as prompt to the text-to-image generation model, which is motivated by the results of Study I in the next section that such a prompt helps to generate topical- and emotional-relevant images. 
If users are unsatisfied with the generated images, they can directly click the ``Keywords based'' button (d) to regenerate another four images based on the keywords. 
If users would like to check generated images using the content of their drafted posts as a prompt to the model, they can click the ``Content based'' button to check four generated images. 
In our experimental setup, the backend server is hosted on a desktop with a GeForce RTX 3070 Ti graphics card. 
Every round of generating four images takes about 5-7 seconds. 
To mitigate the negative impact of delayed response, the buttons ``Generate images'', ``Content based'', and ``Keywords based'' will turn to ``Waiting for about 6 seconds'' when the server is generating the images. 

}

\peng{
\subsubsection{Text-to-Image Generation Model and Prompts}
\label{sec:prompts}
We chose the open-source pre-trained anything-v5 Stable Diffusion Model \footnote{\url{https://stablediffusionapi.com/models/anything-v5}. It has generated over 600K generated images up to December 2023.} for text-to-image generation.
This model is the latest version of the anything series model that generates cartoon-style images, which could mitigate the risk of the model inadvertently infringing on individuals' likeness rights or causing negative psychological experiences for users \cite{Berking12}, e.g., the ``uncanny valley'' effect \cite{Mori12}. 
We did not use the other large models like DALL·E for image generation as it is not freely available and is not accessible in some regions. 
As described in \autoref{sec:user_interface} above, MentalImager supports image generation based on the content of the post or extracted / user-initiated keywords (\autoref{fig:interface}c). 
Correspondingly, the image generation model allows two types of text formats as input, which are also shown in \autoref{fig:methods}: 

\textbf{Content-based}: A descriptive paragraph. We take the textual content of the post as such an input to the model (example in the text area of \autoref{fig:interface}a). 

\textbf{Keyword-based}: Keywords-Weight pairs. We take extracted keywords (example in \autoref{fig:interface}c) from the post as such an input to the model.
we choose the nltk-rake (Rapid Automatic Keyword Extraction) \footnote{\url{https://pypi.org/project/rake-nltk/}} package, a commonly used and powerful approach, for keyword extraction. 
We take two steps using WordNet to make the keywords more understandable by the model, as using the model's known vocabulary would be beneficial for the output \cite{Fast15}. 
First, we remove the pronouns and auxiliary words in the extracted keywords.  
Second, we conduct synonym conversion on the extracted keywords if they have a simpler synonym, e.g., ``enjoy, joy, cheerful, grateful'' are converted to ``happy''. 
We enable users to adjust the weight of each keyword, where a larger number indicates that the keyword plays a more important role in the image generation process. 
\zhenhui{We assign the initial weights to the detected keywords from the posts based on trials and errors on multiple example mental health self-disclosing posts selected from Reddit r/depression. As a default setting, the weights are ``1''. For a keyword that has a top-5 TF-IDF score in the extracted keywords and is a noun or an adjective, we set its initial weight as ``1.1'' as we observe that these words could help generate images matched with the topic of the textual posts.}

The two textual prompts above do not include the title of the post to handle the cases where users have not drafted the title before invoking MentalImager. 
However, based on our trials on several posts, we observe that the images using a keyword-based prompt described above may not be topically and emotionally relevant to the post. 
Therefore, we explore a refined version of the keyword-based prompt: 

\textbf{Emo-keyword-based}: the extracted keywords and detected emotional keyword (e.g., ``anger'' in \autoref{fig:interface}c) from the content of the post + one keyword extracted from the post's title. We use a pre-trained BERT-based model \cite{Hartmann22} to detect the post's perceived emotion, \ie anger, disgust, fear, joy, neutral, sadness, or surprise.  
\zhenhui{
We choose the six classes of emotions for two reasons. 
First, they are basic and easy to understand, and the corresponding keywords can be input to the stable diffusion model for image generation. 
Second, users can adjust the weight of the detected emotional keyword to guide the model to pay less or more attention on that emotion during image generation. 
This adjustment can be viewed as shifting the emotion to another emotion in an arousal-valence map, \eg increasing the weight of ``sadness'' could express an emotion closer to ``depression''. 
Besides, the extracted keywords may also reveal users' emotions (\eg ``happy'' and ``depressed'' in \autoref{fig:post}) if the post mentions them. 
In other words, the key-value pairs of the six emotions and the keyword extraction and edition approach in \name{} could support users to reveal  any emotion that guides the image generator.
Apart from the emotional keyword, we use nltk-rake package to extract a keyword from the title if the post contain it. 
For the emotional keyword and the keyword extracted from the title, we set their initial weight as ``1.3'' as we observed in our trials and errors that this weight could guide the model to generate images that reveal the emotion and topic specified by the keywords. Nevertheless, users can adjust the weights of any keywords as they wish. 
}

\zhenhui{\autoref{fig:methods} shows how images are obtained through the three image generation approaches and the retrieval-based baseline approach in Study I.}

\begin{figure}[h]
  \centering
  \includegraphics[width=\textwidth]{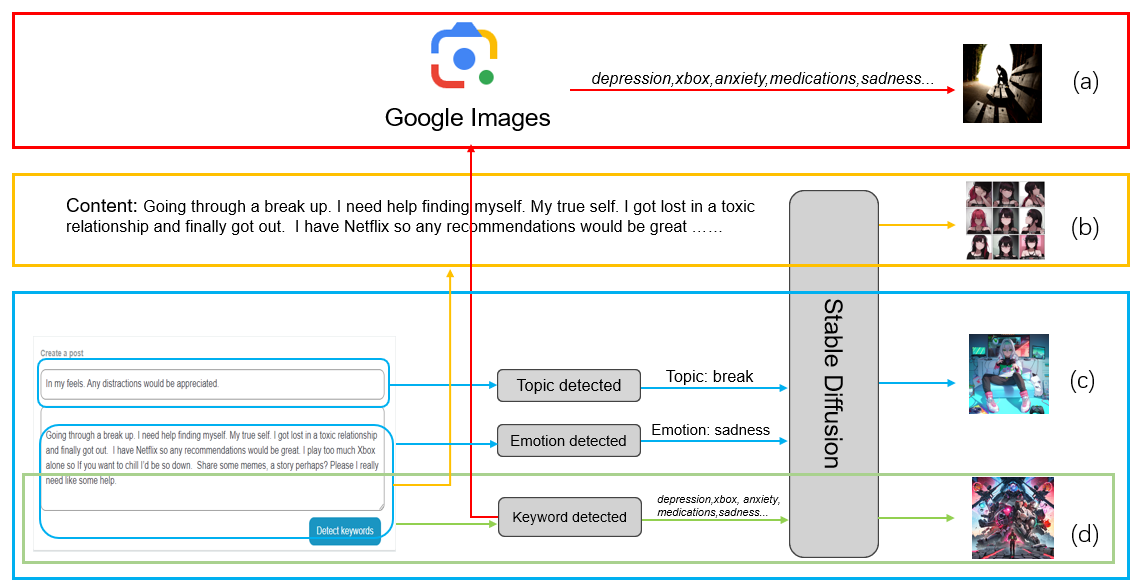} 
  \caption{
  \zhenhui{
  Four methods to generate images, which are:  
  (a) Baseline: Extract keywords from the user text and input them into Google Images for retrieval, using the first downloadable image as the result. In the example, the output is an image of a man sitting on a bench \textit{(red box)}.
  (b) Content-based: Directly input the user text content into the Stable Diffusion (SD) model to generate an image. In the example, the output is a collection of sad face \textit{(orange box)}.
  \textbf{(c) Emo-keyword-based (proposed)}: Input the extracted keywords along with additional topic and emotion keywords into the SD model to generate an image. In the example, the output is a girl in her room with many gaming devices \textit{(blue box)}.
  (d) Keyword-based: Input the extracted keywords into the SD model to generate an image. In the example, the output is a Netflix-style poster image \textit{(green box)}.
  }
  }
  \label{fig:methods}
\end{figure}

}

\peng{
\section{Study I: Relevance of Generated Images to Mental Health Self-Disclosing Posts}
\label{sec:study_1}

As identified by the survey respondents \autoref{sec:survey}, users need MentalImager to generate topical- and emotional-relevant images based on the user's drafted post or specific keywords. 
In this section, we present a human evaluation study to address RQ1: whether and how text-to-image generative techniques can generate images that are topic- and emotion-relevant to support-seeking posts in online mental health communities (OMHCs). 
Specifically, we compare the images generated by the content-based, keyword-based, and emo-keyword-based prompts (\autoref{sec:prompts}) to a baseline approach that retrieves existing images online. 
}

    
    


\peng{
\subsection{Baseline approach for getting relevant images}
To examine whether the generated images can achieve comparable performance with existing images in revealing posters' thoughts and feelings, we implemented a baseline approach for getting relevant images for a post. 
This approach extracts keywords in a similar way to the keyword-based prompts described in \autoref{sec:prompts} and uses the top-5 keywords with the largest TF-IDF scores as input to the Google Image \rl{Search} Engine \footnote{\url{https://www.google.com/imghp?hl=en}}. 
We use the first returned image for the comparison in Study I. 
This approach simulates how a poster can traditionally search for a relevant online image and attach it to their posts. 

\zhenhui{
\subsection{Sampled Posts}
We sampled 20 posts from 69,998 posts created between January 1st, 2022, and May 31st, 2022, and collected via Pushshift API from the Reddit r/depression community. 
This community is a supportive space for anyone struggling with depression and has been widely used as a research site for analyzing the posts and comments in online mental health communities \cite{Peng20,Peng21,sharma2018mental}. 
Two authors of this submission randomly sampled one post from the dataset and checked if it revealed different content \footnote{We observed that many posts in this community are expressing their depression and anxiety without describing the context. We excluded several such posts in this sampling process to include more chosen posts that describe different contexts.} from the previously chosen posts, and if yes, put it in the set of chosen posts and repeat this process until 20 posts are chosen. 
On average, the word count of the 20 posts is 810 (SD = 681). 

We validated the appropriateness of the number and diversity of the randomly sampled posts as below. 
First, we used the G* power 3.1 software to conduct a power analysis on the sample size for Study I. 
Specifically, in our intended one-way repeated-measured ANOVA to assess the differences among the ratings of the four groups of images, we input to G* power with Effect size = 0.40 (large effect), Power $(1 - \beta \text{ err prob}) = 0.8$ (an acceptable threshold), $\alpha \text{ err prob} = 0.05$ (default), Number of groups = 4 (our case), Number of measurements = 3 (our case), Corr among rep measure = 0.5 (default), and Nonsphericity correction $\epsilon = 1$ (default). 
This outputs that the recommended smallest sample size is 16. The 20 posts used in Study I satisfy this minimum requirement.

Second, we examine the diversity of the chosen posts regarding their topics. Specifically, we use Latent Dirichlet Allocation (LDA) to identify and extract abstract topics from the 69,998 posts. 
Following \citet{Li_Wu_Liu_Zhang_Guo_Peng_2024}, 
the topic modeling involves four steps. 
1) Cleaning: we preprocess the titles and text of all the posts by i) removing non-English characters such as emoticons and digits, ii) converting letters into lowercase, iii) removing stop words~(\eg about, the, me), iv) performing stemming for words using the WordNet Lemmatizer from NLTK (\url{https://www.nltk.org/}), v) removing words whose length is shorter than 2 or longer than 15, and vi) removing words whose frequency is less than 5 in our corpus.
2) Building: we use the Gensim 3.8 software to build LDA models by setting the metadata parameters ``the number of passes'' to 100, $\alpha$ to auto, and $\eta$ to auto, which allow the models to infer the asymmetric topic distribution from the corpora. 
3) Evaluation: we use the generated dictionaries and corpora to build 19 separate LDA models with a number of themes ranging from 2 to 20. 
We select the model with the lowest perplexity score, which has 4 topics. 
4) Analysis: we make sense of the clustered topics by analyzing the most frequently appeared words. 
Two authors of this paper had multiple rounds of open-coding and discussions on the names and definitions of these clustered topics.
The four topics are named as ``\textbf{mental health}'' ($34.6\%$)  (0.021$*$`depression' + 0.015$*$`time' + 0.012$*$`people' + 0.010$*$`life' + 0.009$*$`anxiety' + 0.007$*$`mental' + 0.007$*$`school' + 0.007$*$`feeling'), 
``\textbf{emotion}'' ($22.6\%$) (0.026$*$`fucking' + 0.015$*$`hate' + 0.014$*$`shit' + 0.013$*$`fuck' + 0.013$*$`sleep' + 0.009$*$`time' + 0.008$*$`bed' + 0.007$*$`eat'), 
``\textbf{life}'' ($11.4\%$) (0.038$*$`life' + 0.022$*$`people' + 0.013$*$`happy' + 0.012$*$`feeling' + 0.012$*$`time' + 0.010$*$`love' + 0.009$*$`hate' + 0.009$*$`pain'), and
``\textbf{friends or family}'' ($31.4\%$)  (0.017$*$`friends' + 0.013$*$`time' + 0.010$*$`life' + 0.010$*$ `family' + 0.010$*$`told' + 0.010$*$`job' + 0.010$*$`friend' + 0.008$*$`talk').
Two authors of our research team carefully 
Among our selected 20 posts, the primary topic of nine posts is ``emotion'', seven posts are primarily about ``friends or family'', two posts are about ``mental health'', and the rest two posts mainly talk about ``life''. 
The sampled 20 posts with masks on sensitive information for privacy concerns are attached in the supplementary materials. 
This suggests that the chosen 20 posts cover those expressing emotions and those describing their diverse situations. 
}

\subsection{Procedure}

\autoref{fig:post} shows an example of the selected post, its retrieved image, and its three generated images. 
\zhenhui{In total, we have 80 (20 posts \* 4 conditions) images for the human evaluation study.}
We prepared a document that, one by one, lists the post, its four related images in a shuffled order, and the rating scheme for each image. 
The rating scheme includes \textbf{visual quality} (i.e., overall quality of the generated images), \textbf{topic-relevance} (i.e., the degree of relevance of the image's topic to that of the post), and \textbf{emotion-relevance} (i.e., the degree of relevance of the image's emotion to that of the post). 
All items are rated on a 5-point Likert Scale, with 1 being the worst and 5 being the best. 
The visual quality and topic-relevance are adapted from fidelity and alignment in \cite{Otani_2023_CVPR}, while the emotion-relevance metric is specific in our scenario and designed in a similar form to the topic-relevance. 
\taewook{We invited 10 raters (six males, four females, age $M=22.00, SD=1.63$), six of them are university students, one public servant, one community worker, one social worker, and one freelancer}, who reported having experience in using images in their mental health self-disclosure, to rate the images independently and averaged their scores for each image as the final scores.


\begin{figure}[H]
  \centering
  \includegraphics[width=0.6\textwidth]{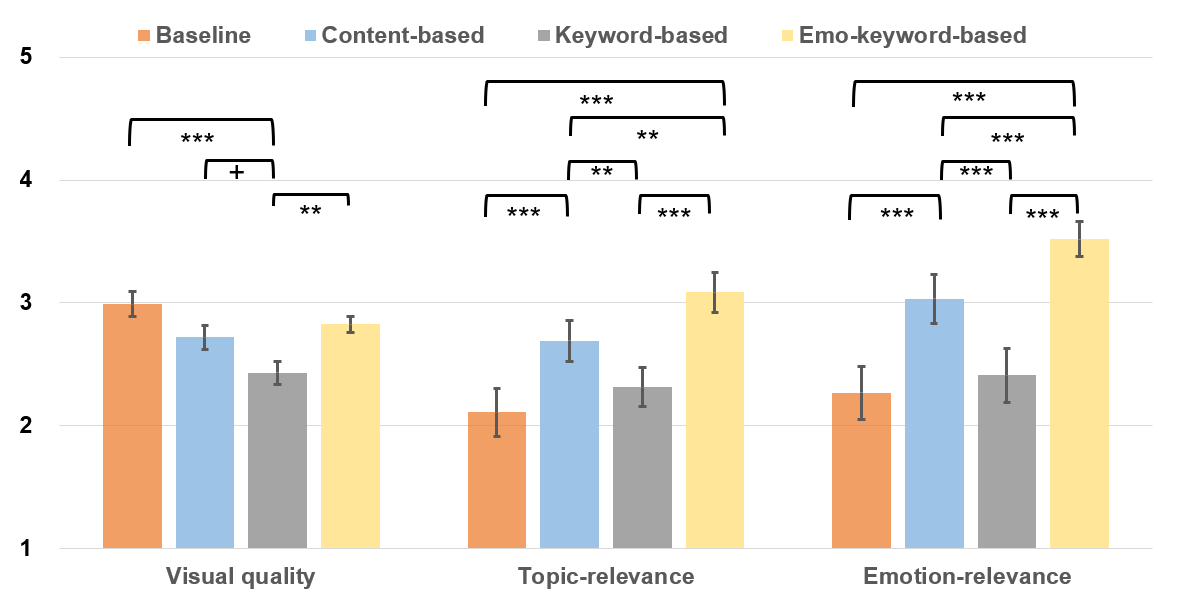} 
  \caption{Means and standard errors of the perceived visual quality of the images as well as their relevance to the input textual posts in an online mental health community. The images are either retrieved using the Baseline approach or generated using Content-based, Keyword-based, or Emo-keyword-based textual prompts. 5-point Likert scale; $+ : .05 < p < .1, * : p < .05, ** : p < .01, *** : p < 0.001$.}
  \label{fig:study_1}
\end{figure}

\begin{figure}[h]
  \centering
  \includegraphics[width=\textwidth]{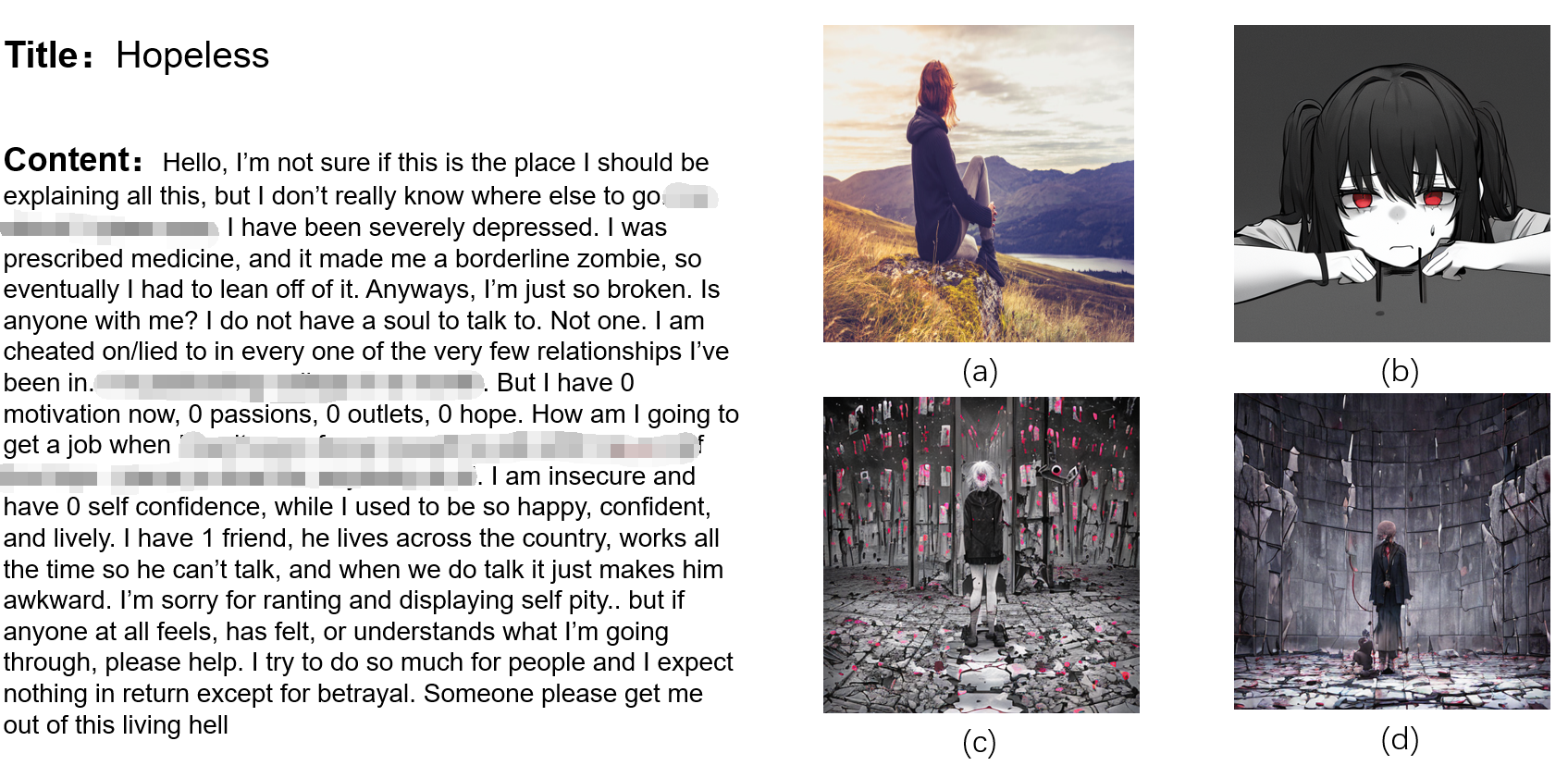} 
  \caption{Example post and its four related images in Study I. (a) Baseline. (b) Content-based. (c) Keyword-based: in this example, the keywords are \textit{``broken,  happy, depressed, pity, displaying self-pity, 0 self-confidence, someone please get, said, works, understands''}. (d) Emo-keyword-based: in this example, apart from the keywords in (c), it additionally uses detected use the emotional keyword \textit{``sadness''} and the keyword from the title \textit{``hopeless''}. We mask sensitive details of the example post for privacy concerns.  
}
  \label{fig:post}
\end{figure}

\zhenhui{
\subsection{Analyses and Results}}
We used one-way repeated-measured ANOVA to assess the statistical differences among the ratings of the four groups of images.
The assumptions for conducting the ANOVA, normal residuals and equal variance, were satisfied. 
For the differences between each two groups of images, we used the Bonferroni post-hoc test. 

\zhenhui{
The results show significant differences among the four image groups in terms of visual quality ($F(3,796)=7.682, p < 0.001, \eta^2=.03$), topic-relevance ($F(3,796)=8.101, p < 0.001, \eta^2=.10$), and emotion-relevance ($F(3,76)=9.93, p < 0.001, \eta^2=.15$). 
Specifically, as shown in \autoref{fig:study_1}, the visual quality of the retrieved images in the baseline ($M=2.99, SD=1.24$) is significantly higher than that of the images generated by keyword-based ($M=2.43, SD=1.17, p < 0.001$), and the same phenomenon was shown between keyword-based and emo-keyword-based prompts ($M=2.83, SD=1.11, p < 0.01$).
This indicates that generating images solely based on keywords extracted from user text may not result in ideal visual quality, while improvements can be observed when additional emo- and topic- keywords are incorporated.
Additionally, in topic-relevance / emotion-relevance, the images generated using content-based prompts ($M=2.69, SD=1.20 / M = 3.04, SD = 1.29$) and emo-keyword-based($M=3.09, SD=1.15 / M = 3.52, SD = 1.10$) have significantly better scores than the retrieved baseline images ($M=2.11, SD=1.11, p < 0.001 / M=2.27, SD=1.21, p < 0.001$).
Besides, the images generated using the emo-keyword-based prompts are significantly more topic-relevant / emotion-relevant ($M=3.09, SD=1.15 / M = 3.52, SD = 1.10$) to the posts than the images using the keyword-based prompts ($M=2.32, SD=1.09, p < 0.001 /  M=2.41, SD=1.23, p < 0.001$) and the baseline images ($p < 0.001 / p < 0.001$).
}

These results indicate that compared to searching existing images online, our adopted anything-v5 stable diffusion model could use content-based and keyword-based textual prompts to generate images that are more topic- and emotion-relevant to the posts in online mental health communities. 
When comparing the images generated using keyword-based and emo-keyword-based prompts, we highlight the importance of considering the keyword of a post's title and the perceived emotion of a post's content in the image generation process. 
Based on these results, we use the emo-keyword-based prompt as a default setting (i.e., when clicking the ``Generate images'' button) for generating images in MentalImager.
We also allow users to directly use the content of the post by clicking the ``Content-based'' button to generate needed images. 

}





\peng{
\zhenhui{
\section{Study II: Effects of MentalImager on Mental Health Self-Disclosure}}
After validating the feasibility of our model for generating relevant images to posts in online mental health communities (OMHCs), we conducted a within-subjects study with 24 participants to address 
RQ2: how would the \zhenhui{MentalImager} impact the outcome and experience of self-disclosure via the posts in OMHCs.

\subsection{Baseline Interface}
The goal of Study II is to evaluate the effects of \zhenhui{MentalImager} on the mental health self-disclosure process and outcome. 
\zhenhui{The key features of MentalImager are the generated images and editable keywords for image generation.}
\zhenhui{Therefore, except for these two features about generative images, the baseline interface should have other features (\eg allow create a post with text and image) of MentalImager to minimize potential confounding factors.}
The buttons for generating images, continuing to submit, or re-editing the post (\autoref{fig:interface}b, b') in MentalImager are replaced by buttons for uploading a local image and submitting the post in the baseline interface (f). 
The baseline interface does not have the keyword editor and image-displaying panels as does in the MentalImager interface. 

\subsection{Participants}\label{sec:participants}
We distributed our recruitment survey on social media at the local university and via word-of-mouth. 
In the survey, we first informed that the experiment would involve the disclosure of personal negative emotions or experiences and obtained consent from the participants. 
We only included the participants who reported encountering mentally challenging issues within two weeks and indicated willingness ( > 4 on a 7-point Likert Scale, 1 - not willing at all, 7 - highly willing) to disclose them in our simulated OMHC.  
We finally included 24 participants (P1-24, 16 males, 8 females, age: $M = 20.9, SD = 1.6$) after screening the first 51 respondents to the survey. 
We conducted a power analysis using the G*power software on the sample size. 
Specifically, in the Wilcoxon signed-rank tests for the measures described in \autoref{sec:measure}, we input to G*power an effect size of 1 (expected means difference of measures: 1, SD: 1), a power of 0.8, and a p-value of 0.05. This outputs that the recommended smallest sample size is 9. 
All participants were undergraduate or master students who have passed the CET-6 (College English Test Band 6) in China, indicating that they have reached the English level of non-English major postgraduates. 
Participants' self-assessment on the Patient Health Questionnaire-9 (PHQ-9) \cite{phq9} showed that 11 of them had mild depression (scores between 5-9), 8 had moderate depression (10-14), and 5 had severe depression (15-27). 
They reported stress associated with exam preparation, anxiety about not achieving the expected learning outcomes, irregular daily routines, a lack of interest in daily life, and so on.


}

\peng{
\subsection{Task and Procedure}
\begin{figure}[h]
  \centering
  \includegraphics[width=\textwidth]{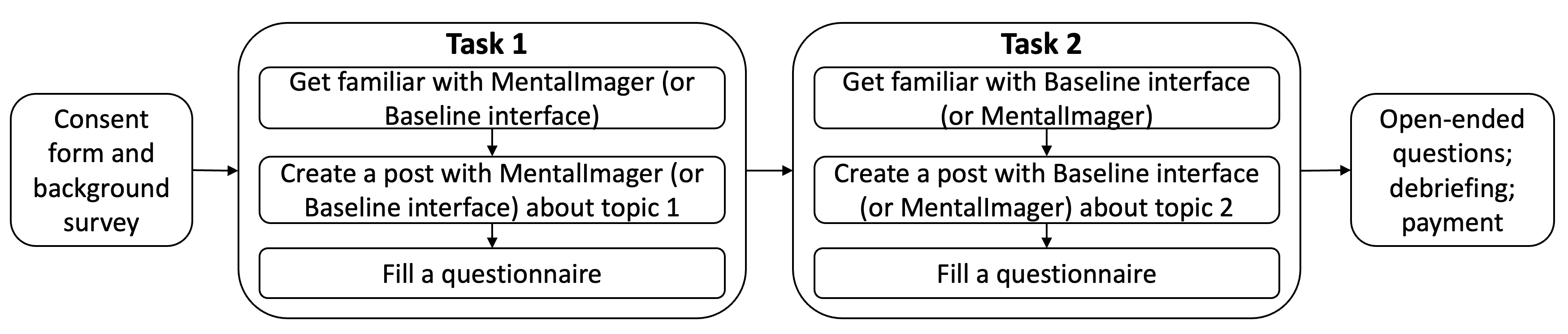} 
  \caption{Procedure of Study II. Participants use our proposed MentalImager interface and the baseline interface in a counterbalanced order to create two posts about different topics they specify in their background survey.} 
  \label{fig:procedure}
\end{figure}
We conducted the study with each participant online with a document that details the task and procedure.  
As shown in \autoref{fig:procedure}, after filling out a consent form and a background survey, participants start their two tasks. 
In each task, participants were required to create a support-seeking post in our simulated OMHC with either MentalImager or the baseline interface. 
The topics of their posts are specified by themselves in the background survey. 
After participants submitted their posts in each task, they filled out a questionnaire about their experience and perception of the used interface. 
We counterbalanced the orders of used interfaces to mitigate the potential order effects. 
Upon completion of the two tasks, participants filled the last questionnaire with open-ended questions about the perceived strengths and weaknesses of MentalImager and suggestions for improvement.  
They then contacted our research team for a debriefing session and got 30 RMB (about 4.2 USD) for compensation of about 30 minutes spent in the study. 
}

\peng{
\subsection{Measures}\label{sec:measure}
\subsubsection{Outcome of mental health self-disclosure}
We capture how the generated images of MentalImager impact the self-disclosure outcome in the posts from the aspects of both support-seekers and support-providers. 
In the questionnaire after each task, as support-seekers, participants rated their perceived \textbf{satisfaction of self-disclosure} in their posts (i.e., ``My mental health disclosure in the post is clear, and I am satisfied with it'', adapted from \cite{Peng20}) in a 7-point Likert Scale; 1 - strongly disagree, 7 - strongly agree.

After the completion of the user study, we invited another six people (3 males, 3 females, age: 20-22, compensation: 30 RMB, about 4.2 USD for 30 minutes) who had experience in posting or commenting in OMHCs to act as the support-providers. 
These six support-providers gave scores (1 - the lowest, 7 - the highest) on each of the shuffled support-seeking posts from our participants regarding two items adapted from \cite{Wang16}: 
\begin{itemize}
\item \textbf{Degree of empathy}: To what extent can you sense the emotions or needs of the poster in this post?
\item \textbf{Willingness to reply}: Would you be willing to provide support or reply to this post?
\end{itemize}
We average the scores of the six raters on each dimension as the final score on that dimension for each post. 

\subsubsection{Experience of mental health self-disclosure}
In the questionnaire after each task, participants rated their \textbf{perceived difficulty} in disclosing their mental health experience, thoughts, and feelings; 1 - very easy, 7 - very difficult. 
In the questionnaire after the MentalImager condition, we also asked participants to share whether and how MentalImager influenced their self-disclosure.  

\subsubsection {Perceptions towards MentalImager}\label{subsubsec:perception}
In the questionnaire after the task with MentalImager, we asked participants to rate the perceived usefulness of MentalImager's generated images for facilitating self-disclosure (four items adapted from \cite{Peng20}, Cronbach $\alpha = .612$, e.g., ``the generated images are useful for me to express my thoughts and emotions in the post''), easiness to use of the interface (four items \cite{Peng20}, $\alpha = .712$, e.g., ``The process of generating and regenerating images is straightforward''), and intention to use (two items \cite{Peng20}, $\alpha = .844$). 
To complement Study I's results about the relevance of generated images to the text, we also asked them to rate their perceived topic-relevance and emotion-relevance between the generated images and text of the post (\ie to what extent do you think the generated images are relevant to the topic / emotion of the textual content in your post?); 1 - least relevant, 7 - most relevant. 
Besides, from support-providers' perspectives, we asked our six human raters to give scores (1 - the lowest, 7 - the highest) on the shuffled support-seeking posts from our participants regarding ``Image-text relevance'' (\ie Do you think the images in the post are relevant to the topic and emotions in the text?). 
This metric is only applicable to the posts that attach an image.

\subsection{Safety Protocol} \label{sec:safety}
We took several approaches to protect participants' privacy and safety in our study. 
First, prior to this work, we obtained IRB approval for a broader research project on patients' and caregivers' practices of healthcare service systems and online communities, covering both our data collection and analysis as well as interviews with participants.
Second, we anonymized participants' information using numbers and masked the sensitive information of the posts when presented to the raters. 
Third, we informed participants that they can exit the experiment at any time they wanted. 
Fourth, we kept in touch with participants via email every day for a week and recommended them to seek professional help if they feel uncomfortable. 
}

\peng{
\subsection{Analyses and Results}
\label{sec: useful}
We performed Wilcoxon signed-rank tests to assess the difference between the MentalImager and baseline conditions regarding participants' perceived satisfaction of self-disclosure, support-providers' degree of empathy on and willingness to reply to the posts, and perceived difficulty of mental health self-disclosure. 
The Wilcoxon signed-rank test is a non-parametric statistical test commonly used in HCI studies  (e.g., \cite{peng2024cscw,wilcoxonuse1}) to compare two matched samples especially when the data do not meet the assumptions of normal distribution. 
For the participants' comments and suggestions on MentalImager, two authors conducted a thematic analysis \cite{thematic}.
They first got familiar with and assigned codes to the participants'  answers to the open-ended questions independently.
After several rounds of coding with comparison and discussion, they finalized the codes of all the responses regarding the pros and cons of MentalImager. 
We incorporated these findings into the following presentation of results.
}

\begin{table}[]
\caption{The Study II's statistical results about outcome and experience of mental health self-disclosure as well as their perceptions towards the interfaces in the MentalImager and the baseline conditions. All items are measured in 7-point Likert scale (1 - the worst; 7 - the best). Note: $* : p < .05, ** : p < .01, *** : p < .001$; Wilcoxon signed-rank test; within-subjects; $N = 24$.}
\label{table_results}
\scalebox{0.9}{
\begin{tabular}{lllllll}
\hline
\multirow{2}{*}{}       & \multirow{2}{*}{Items}                      & \multirow{2}{*}{\begin{tabular}[c]{@{}l@{}}Baseline\\ Mean (SD)\end{tabular}} & \multirow{2}{*}{\begin{tabular}[c]{@{}l@{}}MentalImager\\ Mean (SD)\end{tabular}} & \multicolumn{3}{c}{Statics}                                              \\ \cline{5-7} 
                            &                                             &                                                                               &                                                                                   & \multicolumn{1}{c}{Z} & \multicolumn{1}{c}{p} & \multicolumn{1}{c}{Sig.} \\ \hline
\multirow{3}{*}{Outcome}    & Seeker's satisfaction of the post           & 4.67 (0.87)                                                                   & 5.96 (1.08)                                                                       & -3.332                & .001                  & **                       \\ \cline{2-7} 
                            & Viewers' empathy on the post              & 4.39 (1.43)                                                                   & 5.19 (1.20)                                                                       & -5.695                & .000                  & ***                      \\
                            & Viewers' willingness to reply to the post & 3.93 (1.51)                                                                   & 4.67 (1.47)                                                                       & -4.441                & .000                  & ***                      \\ \hline
Experience                  & Perceived difficulty of self-disclosure     & 3.42 (1.24)                                                                   & 2.62 (1.09)                                                                       & -2.249                & .024                  & *                        \\ \hline
\multirow{6}{*}{Perception} & Usefulness                                  & \multirow{3}{*}{NA}                                                           & 5.54 (1.12)                                                                         & \multicolumn{3}{l}{\multirow{6}{*}{Not Avaliable}}                       \\
                            & Ease of use                                 &                                                                               & 5.51 (1.18)                                                                         & \multicolumn{3}{l}{}                                                     \\
                            & Intention to use                            &                                                                               & 5.75 (0.90)                                                                         & \multicolumn{3}{l}{}                                                     \\ \cline{2-4}
                            & Text-image topic relevance                  & \multirow{2}{*}{NA}                                                           & 5.45 (1.21)                                                                       & \multicolumn{3}{l}{}                                                     \\
                            & Text-image emotion relevance                &                                                                               & 5.20 (1.35)                                                                       & \multicolumn{3}{l}{}                                                     \\ \cline{2-4}
                            & Image-text relevance (viewers' aspect)    & 4.11 (1.70)                                                                    & 4.38 (1.52)                                                                       & \multicolumn{3}{l}{}                                                     \\ \hline
\end{tabular}
}
\end{table}

\peng{
\subsubsection{Outcome of mental health self-disclosure}
In total, with MentalImager, all 24 participants chose to attach generated images to their submitted posts, while with the baseline interface, 16 out of the 24 participants attached images (either from local devices or online resources) to the posts. 
As shown in Table \ref{table_results}, as support-seekers, participants felt significantly more satisfied with their self-disclosure in the posts drafted with MentalImager ($Median = 6.00$) than with the baseline interface ($Mdn = 5.00$); $Z = -3.332, p < 0.01$. 
From the support-providers' perspectives, our human raters felt that the posts written with MentalImager ($Mdn = 5.00$) invoke significantly more empathy than those written with the baseline interface ($Mdn = 4.00$); $Z = -5.938, p < 0.01$. 

Our raters were also significantly more willing to reply to the posts from the MentalImager's condition ($Mdn = 4.00$) than those from the baseline condition ($Mdn = 3.00$); $Z = -4.770, p < 0.01$. 
These results suggest that MentalImager can improve users’ satisfaction with their mental health self-disclosing posts and help the posts invoke support-providers' empathy and willingness to reply.

\begin{figure}[h]

    \includegraphics[width=\linewidth]{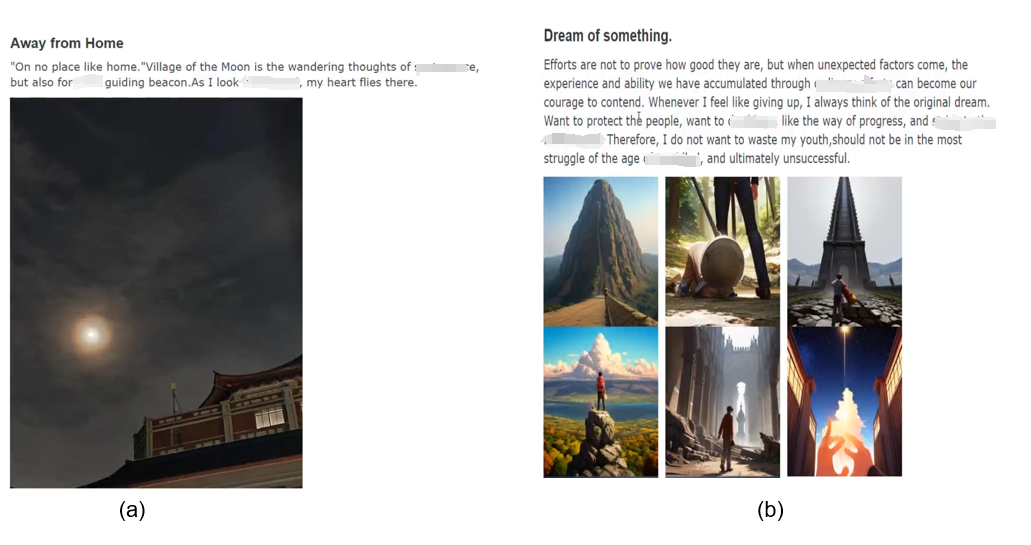}
    \caption{\zhenhui{\textbf{(a)}  P10's post in the baseline condition and \textbf{(b)} P10's post in the MentalImager condition.}}
    \label{fig:user_case1}
\end{figure}

\zhenhui{
\autoref{fig:user_case1} shows the two posts created by P10 (male, age: 20) in Study II. 
In the baseline condition, P10's post is about missing home and includes an image of the moon, which is a symbol of homesickness in Chinese culture. 
In the MentalImager condition, P10 shared a story about being inspired by dreams during difficult times. 
Initially, MentalImager focused on the extracted keyword ``dream'', generating somewhat whimsical images, such as a girl under a starry sky and a dreamy moonlit night. 
Later, as the user adjusted the keywords by adding ``struggle'' and ``success'', removing ``uncomfortable things'', and increasing the weight of ``dream'' to 1.6, MentalImager provided images like a modern city street, a youth walking in ancient architecture, and a youth standing on a mountain peak. Then he added a few sentences to his content, ``Therefore, I do not want to waste my youth, should not be in the most struggle of the age of muddled, and ultimately successful'', and clicked content-based button. MentalImager provided another kind of images, which contains hiking and peaks.
P10 selected and attached several generated images in the submitted post. 
}


}

\peng{
\subsubsection{Experience of mental health self-disclosure}
In general (Table \ref{table_results}), participants felt that compared to the MentalImager condition ($Mdn = 2.50$), it was significantly more difficult for them to disclose their mental health concerns and feelings in the baseline condition ($Mdn = 3.00$); $Z = -2.249, p < 0.05$). 
Participants actively shared the impact of MentalImager on their self-disclosure process, which we summarized into the following two categories. 

\textbf{Self-disclosure in text}. 
Fourteen participants reported that they re-edited their posts with inspirations from the generated images or from the process of editing the keywords for regenerating images. 
For example \zhenhui{(\autoref{fig:user_case1})}, P10 (male, age: 20) stated, \textit{``I wanted to express my struggle at the time when I was confused about my future. However, I did not know how to describe my feelings. When I saw a generated image in which a person stands on top of a mountain, it reminded me of the feeling that I was striving for my dream. Whenever I feel like giving up, it is my original dream that helps me move on''}. 
P13 (male, age: 20) mentioned how he modified the wordings in the post when interacting with the generated images: \textit{``When describing the pain caused by the pressure of my study, I initially used 'stressful'. However, during the process of adjusting the keywords to regenerate images, I found an image in which a girl curls up in a corner, which made me think of a 'bottomless pit' (representing endless stress)''.}
Consequently, P13 changed the post's title from \textit{Stressful} to \textit{Tremendous pressure of study} and wrote \textit{``I feel trapped in a bottomless pit of learning''}.
We also found that modifying keywords for regenerating images could be inspiring, which aligns with previous findings in \cite{Shin22}. 
For example, P3 (male, age: 23) stated, \textit{``I wanted to express my desire for freedom. While adjusting the detected keywords of my drafted post, I suddenly thought of a boat which drifts freely''}. 
She added the keyword ``boat'' to regenerated images, wrote down her thoughts related to the boat, and chose an image of a small boat drifting in the sea as an illustration of her post.

\textbf{Self-disclosure in image}. 
The other ten participants did not re-edit the text after checking the generated images. 
However, five participants indicated that they did not need to modify their text, because the generated images already helped them express their thoughts and feelings. 
For example, P2 (male, age: 23) stated, \textit{``When talking about my failure in the graduate entrance exam, I got a generated image of a happy person holding an admission letter and another image in which a person holding a pen and looking frustrated on an exam paper. These two surprising images exactly convey what I want to say''}. 
P14 (female, age: 24) mentioned that she preferred to use images in the self-disclosure rather than a long text, \textit{``Images can catch one's eyes more quickly than text. I found that the generated images could represent the content of the post, so that I did not need to modify the text''}. 
This phenomenon aligns with \cite{Rich02}, which states that users tend to use images to express abstract emotions.


}

\peng{
\subsubsection{Perceptions towards MentalImager}
On average, participants gave high scores to MentalImager's usefulness for facilitating their mental health self-disclosure ($M = 5.54, SD = 1.12$), its easiness of use ($M = 5.51, SD = 1.18$), and their intentions to use ($M = 5.75, SD = 0.90$) it to draft support-seeking posts in online mental health communities (\autoref{table_results}). 
They generally felt that the topics ($M = 5.45, SD = 1.21$) and perceived emotions ($M = 5.20, SD = 1.35$) of generated images are relevant to the textual content of the posts. 
From the support-providers' perspectives, our human raters perceived that the attached images were moderately relevant to the posts' text regarding the topics and emotions (MentalImager: $M = 4.38, SD = 1.52$; baseline: $M = 4.11, SD = 1.70$). 
These results complement Study I's findings and indicate that the generated images are generally relevant to the posters' text. 

Sixteen participants expressed their favor for MentalImager's generated images, as P3 (male, age: 22) stated, \textit{``With MentalImage, I do not need to go online to search pictures, as the recommended ones generally satisfy my requirements on the emotions and content of the images''}. 
Eleven users said that MentalImager encouraged, inspired, and empowered them to manifest their thoughts and feelings. 
For instance, P8 (male, age: 24) said, \textit{``The images drew my desire to express more and gave me inspirations about what to write''}. 
P11 (female, age: 20) added, \textit{``The images would attract more attention from viewers on my post and could visually convey my emotion''}. 
Besides, seven participants appreciated MentalImager's keyword detection and edition features. 
P8 (male, age: 24) commented, \textit{``I can easily increase the weights of certain keywords, which helps me know the focus of my emotions and enables MentalImager to generate more relevant images''}.

However, participants raised several concerns about the generated images in MentalImager. 
For example, six participants felt that the waiting time (currently about 6 seconds) for generating images should be shortened, which could be achieved by a server with more powerful GPUs. 
We summarize the other concerns into the following three categories. 

\textbf{Style of the generated image.} 
Our MentalImager only supports generating images in a cartoon style to avoid infringing on personal or organizational rights. 
However, seven users commented that cartoon-like images were usually not of their interest. 
For example, P1 (male, age: 22) said, \textit{``I prefer generated images related to movies or songs. The cartoon style is too monotonous for me''}.
P10 (male, age: 20) suggested that \textit{``I would expect that MentalImager can allow me to choose if there are human, animals, or scenery in the generated images''}. 
Besides, five users mentioned that the generative images might not work well for extremely negative topics, as commented by P20 (female, age: 19), \textit{``When I was writing a post about depression, MentalImager generated some dark-themed images, which made me feel frightened''.} 
This could be due to the negative keywords of the posts as the input to the text-to-image generative model.
Although the expression of negative emotion can be beneficial to posters \cite{graham2008positives}, our results raise concerns about`` generating images that convey extremely negative emotions. 

\textbf{Relevance and quality of the generated images.} 
Nine participants reported cases where the generated images were somehow irrelevant to the posts, and four commented that the quality of the generated images should be improved. 
P6 (male, age: 20) stated, 
\textit{``I was discussing my negative experience in the post about certain traditional cultures in China. The generated images could not match what I thought''}.
P20 (female, age: 19) added, \textit{``Some characters are blurred. The details of the images can be strengthened.''}
These issues could be due to that our adopted pre-trained text-to-image generative model does not have enough training data that matches certain topics or reveal subtle emotions. 
P17 (male, age: 21) suggested, \textit{``The model should `eat' more data to better handle users' emotions''}.
To improve the quality of generated images for mental health topics, we urge future work to collect mental health self-disclosure text and images to conduct fine-tuning or in-context learning of the text-to-image generative models. 

\textbf{Prominence of the extracted keywords.} 
Five participants reported that some extracted keywords were not the prominent ones in their drafted posts, especially when the posts were still rough or contained multiple topics. 
For example, P21 (female, age: 21) shared that \textit{``In the drafted post, I vent my emotions in a casual and scattered way. The extracted keywords below the post did not correctly reveal my true thoughts''}. 
Using Large Language Models (LLMs, e.g., GPT-4) for keyword extraction \cite{Beeferman23, Purohit23} could help to improve the prominence of the extracted keywords. 
However, it would still be challenging for MentalImager to get the keywords that reveal a user's thoughts and feelings if the user could not clearly state them in the drafted post.

}

\rv{\zhenhui{
\section{Study III: Effects of Generated Images on the Outcome of Mental Health Self-Disclosure}

The findings of Study II have indicated that MentalImager significantly improves users' satisfaction with their support-seeking posts and decreases their difficulties with mental health self-disclosure compared to the baseline community interface without MentalImager. 
However, as the key features of MentalImager include the generated images and editable keywords for image generation, there is a need to separately examine the effects of generated images on mental health self-disclosure. 
To address this need, we conducted a within-subjects Study III with another 30 participants. 
Compared to Study II, the key changes in Study III include participants with more diverse backgrounds, the pre- and post-moment experimental setup, the revised submission flow in MentalImager to satisfy the setup, and the updated measures of self-disclosure outcome after an interview with an expert. 
We detail these changes in the following subsections.

}

\zhenhui{
\subsection{Experimental Setups}
\textbf{Participants}. We invited another 30 participants (noted as P'1 - P'30, 10 females, 20 males, age mean = 23.6, SD = 2.31) in China via word-of-mouth. 
This number also satisfies the recommended smallest sample size (\ie 9) calculated using the same parameters in G*power software, as conducted in Study II (\autoref{sec:participants}). 
Similar to the inclusion criteria in Study II, the participants should report encountering mentally challenging issues within two weeks and indicate a willingness to disclose them in our simulated online mental health communities (OMHCs). 
Among the 30 participants, we have 3 undergraduate students, 3 graduate students, 4 workers in the service industry, 4 office staff, 1 public servant, 5 doctors, and 10 people with other occupations (\eg freelancer, self-employment). 

\textbf{Pre- vs. Post-moment}. 
To evaluate the effects of generated images on the self-disclosure outcome, we follow \citet{Peng20} and define two timestamps in a posting procedure in our study:

\begin{itemize}
    \item \textbf{Pre-}:  the moment right before a user's very first click on the ``Generate Image'' button. At this moment, the user has seen the detected keywords from the draft but has not seen the generated images.
    \item \textbf{Post-}: the moment when a user clicks the  ``Continue to submit'' button  – the user finalizes a post after seeing the generated images.
\end{itemize}

The comparisons of metrics between pre- and post-moments can reveal the effects of the generated images alone, as in the Pre-moment, while users can check and edit the detected keywords, they have not seen the images. 
We do not adopt within-subjects design on the factor of the interface (\eg MentalImager vs. a baseline with editable keywords but without generated images) for two reasons. First, it is somehow repeated in the setup in Study II. Second, as indicated by the results (reported at the end of \autoref{sec:study_1}) of Study I, by default, the generated images are based on the keywords. 
The editable keywords without generated images would serve no purpose and could look weird in the submission flow. 

\textbf{Modified MentalImager}.
To enable us to measure participants' perceptions and outcomes at pre- and post-moments, we modify the submission flow of MentalImager. 
Specifically, users can write the very first draft in the text areas and click the ``Detect keywords'' button (\autoref{fig:interface}b''). 
The interface will then show up the keyword editor panel but \textit{will not} show up the image displaying panel, and the ``Detect keywords'' button will be replaced by a ``Generate images'' button (b'). 
If they wish, users can edit the keywords and the draft of the post at this pre-moment. 
They can then click the ``Generate images'' to show up the image displaying panel for the first time, which will replace the ``Generate images'' button by ``Back to edit'' and ``Continue to submit'' buttons (b). 
Compared to the ``Generate images <-> (Back to edit) -> Continue to submit'' submission flow in Study II (\autoref{sec:user_interface}), such a ``Detect keywords -> Generate images <-> (Back to edit) -> Continue to submit'' flow allows users to check the keywords first and edit them for image generation if they wish, as well as allows us to capture the changes of posts and user perceptions before and after seeing the generated images. 

Apart from the modification in the submission flow, we compiled a word list to exclude extremely negative keywords for image generation in MentalImager, which could mitigate the concerns of generating images that convey extremely negative emotions as reported in Study II. 
Considering that there is no such word list available online \footnote{There are negative vocabulary word lists (\eg \url{https://www.enchantedlearning.com/wordlist/negativewords.shtml}) that contain words like ``confused'' and ``cry'', which, however, are common in mental health self-disclosure and are not extremely negative in our consideration.}, one author of this paper prompted GPT-4o to ``provide 100 extremely negative words about violence, sexual, and disgusting. Please also show the other forms of the generated words''. 
The author went through the generated words and excluded the ones that the author considered as not extremely negative words, which left 307 words as attached in the Appendix. 
If any keyword in the keyword editor panel falls into our compiled word list, MentalImager will automatically exclude these keywords when prompting the image generator. 

\textbf{Task and procedure}.
The task for participants is to create a mental health self-disclosure post in MentalImager about the recent events they found stressful. 
We conducted the study with each participant online with a document that details the task and procedure.
After obtaining participants' consent, we required them to record the screen of their computers while creating the posts in MentalImager. 
After the participants finished the task by hitting the ``Continue to submit'' button, we asked them to fill out a questionnaire that measures their perceptions of the self-disclosure outcome at both pre- and post-moments. 
They then contacted our research team for a debriefing session and got 20 RMB (about 2.8 USD) as compensation for about 20 minutes spent in the study. 

We adopted a similar safety protocol (\autoref{sec:safety}) to that in Study II. 

}

\zhenhui{
\subsection{
Measures
}
The primary goal of Study III is to examine the effects of generated images on the self-disclosure outcome. 
We do not measure the perceived difficulty of mental health self-disclosure as did in Study II, because in Study III, there is no comparison regarding this item between MentalImager and baseline interface. 

\subsubsection{Outcome of mental health self-disclosure}
Before running Study III, we conducted a 30-minute interview with an expert (male, age: 28) who holds a mental health first aid certificate issued by MHFA \footnote{MHFA is an evidence-based, early-intervention course that teaches participants about mental health and substance use challenges (\url{https://www.mentalhealthfirstaid.org/}).}. 
The interview started with introducing the interface of MentalImager and user task in Study III and then focused on seeking the expert's opinions and suggestions on the measures used in Study II (\autoref{sec:measure}). 
The expert raised the following two key points that helped us refine the measures for Study III.
\begin{itemize}
    \item The item ``satisfaction'' (\ie ``My mental health disclosure in the post is clear, and I am satisfied with it'') used in Study II could be more specific in various aspects, \eg clarity of self-disclosure, feeling free to share, and worry about the misunderstanding. 
    \item The item ``degree of empathy'' (\ie ``To what extent can you sense the emotions or needs of the poster in this post?'') asks more about the support-providers' empathy on the posters' feelings, while the providers' understandings of the posters' experiences are also important for supporting the posters. 
\end{itemize}
The expert helped us refine the measures of self-disclosure outcome in Study III as below. 
In the questionnaire after the task, as support-seekers, participants rated their perceptions on their posts at both pre- and post-moments in three items in a 7-point Likert Scale; 1 - strongly disagree, 7 - strongly agree. 
\begin{itemize}
    \item \textbf{Clear self-disclosure}: My mental health disclosure in the post is clear. 
    \item \textbf{Free to share}: I did not hesitate or fear sharing myself in the post. 
    \item \textbf{No worry}: I did not worry that viewers would misunderstand me via the post. 
\end{itemize}
We specified the pre-moment and post-moment in the questionnaire and asked participants to recall these moments when rating their agreement levels of the three items mentioned above at pre-moment and the same three items at post-moment.

As the support-providers, we invited another six viewers (4 males, 2 females, age mean = 23.00, SD = 1.57, compensation: 30 RMB, about 4.2 USD for 30 minutes) who had experience in posting or commenting in OMHCs. 
These six support-providers gave scores (1 - the lowest, 7 - the highest) on each of the shuffled support-seeking posts at both pre- and post-moments from our participants regarding the following items: 
\begin{itemize}
\item \textbf{Degree of empathy}: To what extent can you sense the emotions or needs of the poster in this post?
\item \textbf{Degree of understanding}: To what extent can you understand the experience and encountered problems of the poster in this post?
\item \textbf{Willingness to reply}: Would you be willing to provide support or reply to this post?
\end{itemize}
We average the scores of the six raters on each dimension as the final score on that dimension for each post. 
Specifically, all the 30 participants' posts at pre-moment do not include an image as they only check the detected keywords at this moment, while all the 30 posts at post-moment attach generated images, as observed in our Study III. 


\subsubsection {Perceptions towards MentalImager}
Similar to that in Study II (\autoref{sec:measure}), in the questionnaire, we asked participants to rate the perceived usefulness of MentalImager's generated images for facilitating self-disclosure, easiness to use of the interface, and intention to use MentalImager in their future mental health self-disclosure. 
We also asked them to rate their perceived topic-relevance and emotion-relevance between the generated images and text of the post (1 - least relevant, 7 - most relevant). 
At the end of the questionnaire, we asked participants to explain why they chose the generated images for their posts or why not if they did not choose any generated image. 
Besides, from support-providers' perspectives, we asked our six human raters to give scores (1 - the lowest, 7 - the highest) on the shuffled support-seeking posts from our participants regarding ``Image-text relevance'' (\ie Do you think the images in the post are relevant to the topic and emotions in the text?). 
This metric is only applicable to the posts at the post-moment that attach an image. 
}

\zhenhui{
\subsection{Analyses and Results}
Similar to the analytic approach in Study II, we performed Wilcoxon signed-rank tests to assess the difference between the pre- and post-moments regarding participants' perceptions of their self-disclosure and support-providers' perceptions of the posts in both moments. 
For the participants' responses to why they chose or did not choose the generated images for their posts, two of the authors conducted several rounds of open coding with comparison and discussion. 
We present these qualitative findings in \autoref{sec:perception_study_III}. 
}


\zhenhui{
\subsubsection{Outcome of mental health self-disclosure}
In total, all 30 participants chose to attach generated images to their submitted posts at post-moment. 
As shown in \autoref{tab:study_III}, as support-seekers, participants felt that their mental health self-disclosure is significantly more clear at post-moment ($Mdn = 3.00$) compared to that at pre-moment ($Mdn = 6.00$); $Z=-2.263, p < 0.05$. 
After they had seen and used the generated images at post-moment ($Mdn = 3.00$), they generally felt that they were more free to share themselves in the posts than that at pre-moment ($Mdn = 5.00$); $Z=-1.441, p = 0.150$. 
Furthermore, participants reported that they had significantly less worry about that viewers would misunderstand them via the post at post-moment ($Mdn = 5.00$) than that at pre-moment ($Mdn = 2.00$); $Z=-4.627, p < 0.001$. 

}

\begin{table}[]
\caption{\zhenhui{Study III's statistical results about outcome of mental health self-disclosure at pre- and post-moments as well as user perceptions towards the MentalImager and generated images. All items are measured on a 7-point Likert scale (1 - the worst; 7 - the best). Note: $* : p < .05, ** : p < .01, *** : p < .001$; Wilcoxon signed-rank test; within-subjects; $N = 30$.}}
\label{tab:study_III}
\scalebox{0.9}{
\begin{tabular}{lllcllll}
\cline{2-8}
                            & \multicolumn{2}{c}{\multirow{2}{*}{Items}}             & \multicolumn{2}{c}{\begin{tabular}[c]{@{}c@{}}Moment \\ Mean (SD)\end{tabular}} & \multicolumn{3}{c}{Statics}                                              \\ \cline{4-8} 
                            & \multicolumn{2}{c}{}                                   & Pre-                                      & \multicolumn{1}{c}{Post-}           & \multicolumn{1}{c}{Z} & \multicolumn{1}{c}{p} & \multicolumn{1}{c}{Sig.} \\ \hline
\multirow{6}{*}{Outcome}    & \multirow{3}{*}{Seeker} & Clear self-disclosure        & \multicolumn{1}{l}{4.47 (1.01)}           & 5.10 (1.56)                         & -2.263                & 0.024                 & *                        \\
                            &                         & Free to share                & \multicolumn{1}{l}{5.13 (1.17)}           & 5.53 (1.96)                         & -1.441                & 0.150                 & -                        \\
                            &                         & No worry                     & \multicolumn{1}{l}{3.03 (1.45)}           & 5.90 (0.92)                         & -4.627                & 0.000                 & ***                      \\ \cline{2-8} 
                            & \multirow{3}{*}{Viewer} & Degree of empathy            & \multicolumn{1}{l}{3.72 (1.23)}           & 4.59 (1.37)                         & -8.138                & 0.000                 & ***                      \\
                            &                         & Degree of understanding      & \multicolumn{1}{l}{3.97 (1.41)}           & 4.84 (1.37)                         & -8.090                & 0.000                 & ***                      \\
                            &                         & Willingness to reply         & \multicolumn{1}{l}{3.71 (1.16)}           & 4.22 (1.38)                         & -8.289                & 0.000                 & ***                      \\ \hline
\multirow{6}{*}{Perception} & \multirow{5}{*}{Seeker} & Usefulness                   & \multicolumn{2}{c}{5.41 (1.34)}                                                 & \multicolumn{3}{l}{\multirow{6}{*}{Not Available}}                       \\
                            &                         & Ease of use                  & \multicolumn{2}{c}{5.67 (1.30)}                                                 & \multicolumn{3}{l}{}                                                     \\
                            &                         & Intention to use             & \multicolumn{2}{c}{5.60 (1.36)}                                                 & \multicolumn{3}{l}{}                                                     \\ \cline{3-5}
                            &                         & Text-image topic relevance   & \multicolumn{2}{c}{5.79 (1.13)}                                                 & \multicolumn{3}{l}{}                                                     \\
                            &                         & Text-image emotion relevance & \multicolumn{2}{c}{5.50 (1.43)}                                                 & \multicolumn{3}{l}{}                                                     \\ \cline{2-5}
                            & Viewer                  & Image-text relevance         & \multicolumn{2}{c}{4.62 (1.46)}                                                 & \multicolumn{3}{l}{}                                                     \\ \hline
\end{tabular}
}
\end{table}



\zhenhui{
From the support-providers' perspectives, our human viewers rate that they can significantly better sense the emotions or needs of the poster via the posts with generated images at post-moment ($Mdn = 5.00$) than those without images at pre-moment ($Mdn = 4.00$); $Z=-8.138, p < 0.001$. 
Meanwhile, viewers can significantly better understand the experience and encountered problems of the poster in the posts at post-moment ($Mdn = 5.00$) than those at pre-moment ($Mdn = 4.00$); $Z=-8.090, p < 0.001$. 
Besides, they were also significantly more willing to reply to the posts at post-moment ($Mdn = 5.00$) than those at pre-moment ($Mdn = 4.00$); $Z=-8.289, p < 0.001$. 
These results indicate that the generated images can enhance users' positive feelings about their posts and help the posts invoke support-providers' empathy, understanding, and willingness to reply. 
}

\zhenhui{
\subsubsection{Perceptions towards MentalImager and its generated images} \label{sec:perception_study_III}
In general, participants in Study III gave high scores to MentalImager's usefulness for facilitating their mental health self-disclosure ($M = 5.41, SD = 1.13$), its easiness of use ($M = 5.67,
SD = 1.30$), and their intentions to use ($M = 5.60, SD = 1.36$) it to draft support-seeking posts in online mental health communities (\autoref{tab:study_III}). 
They felt that the topics ($M = 5.79, SD = 1.13$) and perceived emotions ($M = 5.50, SD = 1.43$) of generated images are quite relevant to the textual content of the posts. 
From the viewers' perspectives, our human raters perceived that the generated images were moderately relevant to the posts' text regarding the topics and emotions ($M = 4.62, SD = 1.46$). 
These ratings are similar to those in Study II (\autoref{table_results}) and suggest that the generated images are generally relevant to the posters' text.
We summarize the reasons why participants chose the generated images for their posts below. 
}


\begin{figure}
  \centering
  \includegraphics[width=1\textwidth]{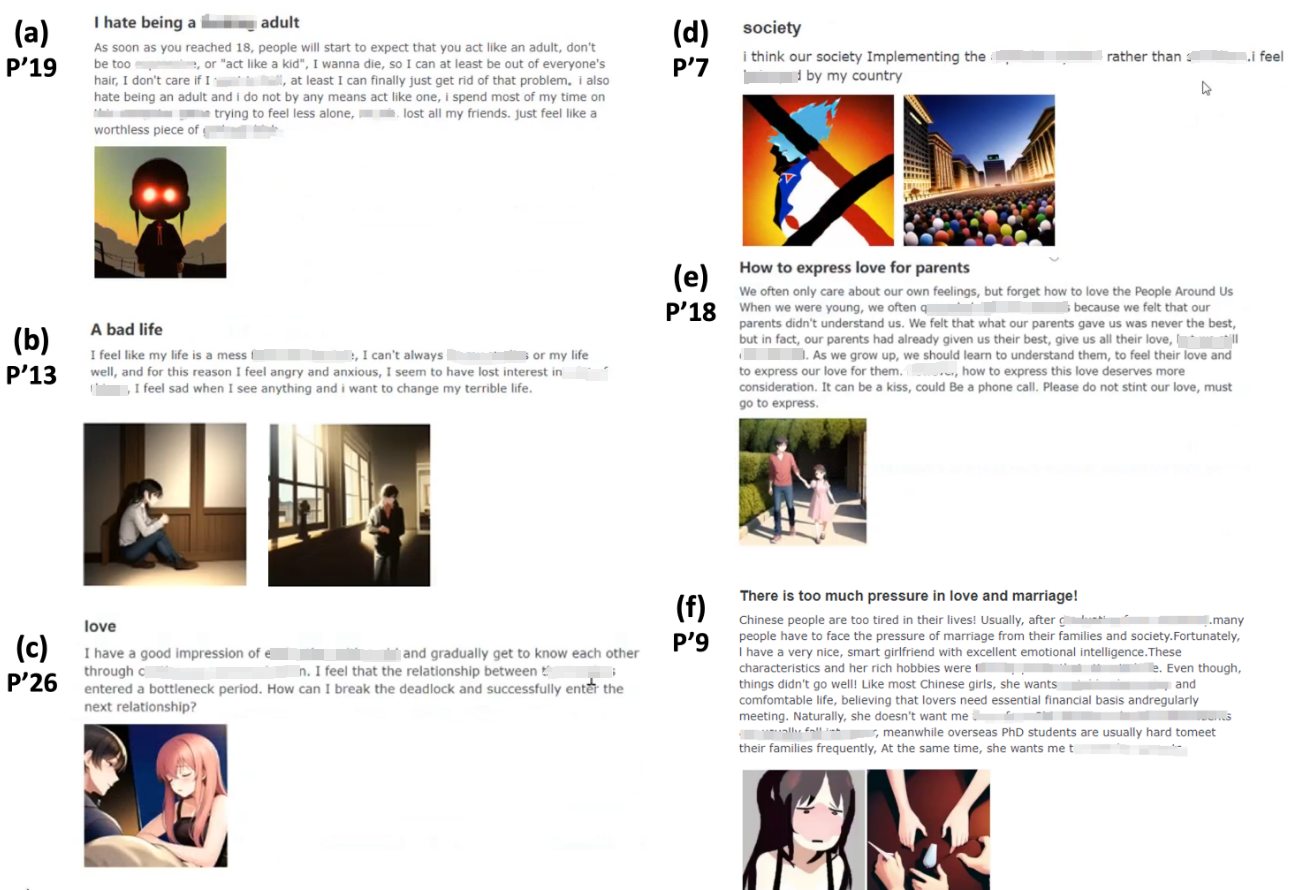} 
  \caption{\zhenhui{Example posts contributed by participants in Study III. We mask some text for privacy concerns.}}
  \label{fig: samples}
\end{figure}

\zhenhui{
\textbf{The generated images help participants express their emotions.}  
Sixteen participants stated that they have chosen to attach generated images in their posts because the images can help them express their emotions. 
P'19 (female, age: 24, self-employment) explained the attached generated image in her submitted post (\autoref{fig: samples}a) at post-moment. 
The post talks about her anxiety about becoming an adult. 
MentalImager identified a set of keywords from the post, including the ``kid'' but excluding the word ``fucking'' and ``hell'' which are in the list of extremely negative words, for image generation. P'19 selected an image of a red-eyed girl and explained, \textit{``This picture looks like an angry little girl, which reflects my current emotional state''}. 

P'13 (female, 25, doctor) described her feeling with the generated images in detail, \textit{``The picture [\autoref{fig: samples}b] I chose resonates with me. It reminds me of when I was a kid, my parents were not around. I spent the day alone, watching the sky getting darker and darker, and finally turning into the color as that picture shows. This picture made me feel lonely. Now my parents are around me, but we have no topics to communicate with as if we are more distant than when we were kids. I think that picture describes my mood very well, so I chose it''}. 
Besides, P'26 (male, 25), who is also a doctor, explained, \textit{``This picture [\autoref{fig: samples}c] depicts a boy looking at a girl affectionately. We can feel that he likes this girl. It fits my current mood, so I chose him''}. 
P'7 (male, 24, postgraduate student) also shared [\autoref{fig: samples} d], \textit{``The first picture implies division, which well reflects my concerns about the current society. The second picture implies that every citizen in society is like a balloon, with a hollow interior and a beautiful appearance, which reflects the impetuous atmosphere of the entire society. It is consistent with my feelings''}. 

\textbf{The generated images represent the content of the text.} 
Nine participants mentioned that the generated images were relevant to the content of the text in their posts. 
P'18 (male, 25, office staff) said, \textit{``[I chosen this image (\autoref{fig: samples}e)] because the father and daughter holding hands in the image have the same scenario as what I want to tell''}. 

\textbf{The generated images can attract viewers' attention and help them understand the posters.} 
Seven participants stated that the generated images can attract viewers' attention as they are \textit{``concise''} (P'30, female, 24, worker) and have a \textit{``good looking''} (P'16, female, 25, public servant).
Two participants suggested that these images can \textit{``help people understand [the poster] better''} (P'4, female, 24, worker). 
}


\zhenhui{
Apart from the support-seekers' perspectives, we encouraged our six human viewers (V1-6) to share their impressed-generated images after the rating process. 
For example, V6 (female, 20, community worker) shared, \textit{``I really liked the image [in P'7's post (\autoref{fig: samples})d]. It looked like a blend of various national flags, and I can feel his interest in the political systems of various countries. Even though it discussed a political topic I disliked, I still understood the message and was willing to discuss it with the poster''}. 
V3 (female, 23, community worker) mentioned, \textit{``I did not have a strong feeling when reading the text version [\ie P'21 's post at pre-moment]. However, the image of a father walking with his child [in the post (\autoref{fig: samples}e) at post-comment] reminded me of my father''}. 
V1 (male, 25, programmer) also shares, 
\textit{``I liked the image of the girl who looked expressionless [in P'9's post (\autoref{fig: samples}f)]. I could feel the poster's mood, filled with helplessness and sadness, just like her face''}.
}



}

\section{Discussion}
\peng{
In this work, we explore the needs, feasibility, and impact of generated images to facilitate support-seekers' self-disclosure in online mental health communities (OMHCs).
In the need-finding survey with 25 respondents, we highlight the user needs for a tool that can interactively generate relevant images to facilitate their self-disclosure in OMHCs, which informs our design of MentalImager. 
In Study I, we validate the feasibility of a state-of-the-art Stable Diffusion model for generating images that are topical- and emotional-relevant to their mental health self-disclosure text.
In Study II \zhenhui{and Study III}, we further show that users can leverage generative images in MentalImager to satisfy their expressive needs in disclosing mentally challenging experiences, thoughts, and feelings. 
These findings extend previous work on images for self-disclosure \cite{Rich02, Kaplan10, park15, Manikonda17} by providing empirical evidence on the benefits of generative images to facilitate self-disclosure of mental health issues. 
Interestingly, participants adjusted the weights of the detected keywords or specified keywords to infuse their emotions into the process of generating images, which, in turn, inspired them to re-edit the textual content in their posts. 
Such an interactive text-image generation process can help users strike a balance between sharing information related to specific needs and the desire to manage self-presentation in social media like OMHCs \cite{Newman11,Luo20}. 
Our findings \zhenhui{in Study II and Study III} also indicate that generative images could enable clearer communication of support-seekers' thoughts and feelings to the support-providers, which validates the power of images for improved communication \cite{draw_pic}. 
We found that the posts created with MentalImager hold the potential to elicit higher empathy and willingness to reply from support-providers, which can improve the posters' social well-being and reduce psychological distress \cite{Luo20}. 
Nevertheless, participants with MentalImager raised several concerns about the style and quality of its generated images. 
These results inform the following design considerations for future intelligent tools that leverage generative images to facilitate self-disclosure in OMHCs. 
}

\peng{
\subsection{Design Considerations for Facilitating Self-Disclosure with Generative Images}
\subsubsection{Conducting fine-tuning or in-context learning of the text-to-image generative models for mental health self-disclosure} 
MentalImager utilizes the anything-v5 Stable Diffusion model to generate images. 
While our Study I validates its feasibility for generating cartoon-like images that are topical- and emotional-relevant to the sampled posts in an OMHC, participants in Study II suggested that the cartoon style may not be of their interests and some images were somehow irrelevant. 
To improve the relevance and usefulness of the generated images, \zhenhui{future work could start from collecting and analyzing seekers' practices and preferences in utilizing images for mental health self-disclosure, \eg styles of images, topical and emotional relationship with the textual disclosure \cite{Li_Wu_Liu_Zhang_Guo_Peng_2024}.} 
For example, we could collect posts that contain both text and images in Reddit's OMHCs (e.g., r/GriefSupport \footnote{\url{https://www.reddit.com/r/GriefSupport/}}) \zhenhui{\cite{Li_Wu_Liu_Zhang_Guo_Peng_2024}}, depression-related imagery and corresponding tags in Instagram \cite{Andalibi15}, and posts in Facebook mental health groups \cite{de2014characterizing,park15}. 
\zhenhui{These posts with identified styles of images and text-image relationships \cite{Li_Wu_Liu_Zhang_Guo_Peng_2024}, on the one hand, can be used to fine-tune the text-image generative models to generate relevant images to mental health self-disclosure text.}
\zhenhui{On the other hand, they can provide examples of images that match to the user-specified style and text-image relationship for in-context learning of the models. }
This would require prompt engineering that describes a task and formulates a set of examples in natural language to a ``prompt'' to the model. 
In either fine-tuning or in-context learning of the generative models, based on our Study I's findings (\autoref{fig:study_1}), we could augment the dataset by adding keywords about the post's perceived emotions.

\subsubsection{Guiding the image generation process with multi-modal prompts}

\rl{MentalImager, a tool for image generation, utilizes either keywords or the full content of a post for its creations. In Study II, while participants favored the keyword-based feature (as shown in \autoref{fig:interface}e), they encountered scenarios where the extracted keywords were not sufficiently descriptive, or their mental health thoughts were too complex to be succinctly conveyed in text for guiding the image generation process. An innovative approach, such as integrating ControlNet \cite{zhang2023adding}, could enhance this process. ControlNet's techniques, like incorporating Canny edges, human poses, and sketches, could offer more intuitive and precise control over image generation. For instance, starting with a sketch could provide a clearer and more direct representation of a user's thoughts, especially in cases where text descriptions fall short of capturing vague thoughts and emotions.}
These multi-modal prompts are promising to address the limitations of extracted keywords that Study II identifies to make it easier for users to leverage generated images for disclosing challenging topics. 

\subsubsection{Involving healthcare elements in the mental health self-disclosure process}
MentalImager helps participants write posts that can invoke significantly more empathy and a higher willingness to reply from support-providers compared to the baseline interface without generated images. 
The replies from others in OMHCs could improve posters' well-being \cite{Luo20}. 
However, many support-seeking posts in OMHCs often can not get timely and needed replies \cite{cass_cscw2021}, which could be due to not only the posts' content but also other reasons like support-seekers' post experience in the community \cite{Peng21}. 
As such, we recommend that tools like MentalImager should provide healthcare content in the process of facilitating mental health self-disclosure. 
For example, they could embed a conversational agent similar to the one in \cite{cass_cscw2021} that provides emotional support after users draft a support-seeking post. 
This conversational agent could also ask questions to guide users to express their thoughts and feelings step by step and conversationally generate needed images.

}

\peng{
\subsection{Implications to the CSCW Community}


Our work offers new insights into AI-mediated communication to the CSCW community. 
Previous HCI and CSCW research (e.g., \cite{Peng20,diyi_modeling_cscw22,Lee20,Shin22}) has mainly focused on developing AI tools that mediate text-based communication between support-seekers and support-providers. 
We focus on using generated images to help seekers express suffering experiences and feelings in their posts. 
We found in Study II that these images can improve seekers' satisfaction with their posts and improve viewers' willingness to provide support to the posts. 
These results indicate the benefit of using generated images to mediate peer support chats. 
More broadly, our results can inform the usage and design of generated images to support communication in other contexts like romantic conversation \cite{Kim:2019} and small talks in instant messaging apps \cite{draw_pic}. 
For example, people can query the AI tools with keywords to generate a romantic image that helps them express their love to others.
Based on our findings in Study I, we highlight the need to infuse users' emotions in the generated images to mediate such communication.


Our work could also shed light on how generated content would affect the community's engagement in the support-seeking posts. 
\citet{cass_cscw2021} developed a conversational agent that acts like a human to provide emotional support to the posts without timely comments and found that the generated comment helps the post attract more responses from the community members. 
While we have not tested MentalImager in a real community due to the ethical concerns \rl{of deploying technology at this early stage of development}, our results about viewers' perceptions of the posts suggest the potential of attaching generated images in the post or comments for engaging community members to interact with the seekers.

Our findings on how the generated images and the process of editing keywords inspire participants' self-disclosure in Study II could support future CSCW research on assisting art therapy or expressive writing with generative technology. 
In an art therapy session, an art therapist would facilitate clients to express their feelings and communicate through the art-making process with structured (e.g., therapists give specific instructions) or unstructured (e.g., clients decide what to draw) directives, followed by a discussion of the artwork, problems and needs \cite{MCNEILLY1983211}. 
\citet{DU2024103139} presents an application named DeepThInk that provides image-to-image generation (e.g., style transfer, brush to photo-realistic images) support in users' art therapy process with or without a therapist. 
We suggest that digital tools could incorporate text-to-image generation techniques like MentalImager to help users express their mental health feelings in the art therapy process. 
As informed by results about user expectation of diverse image styles in Study II and the practices of attaching images for mental health self-disclosure in social media \cite{Burke16, Andalibi15}, we recommend that the generated images should further adapt to users' backgrounds and interests.



}

\subsection{Ethical Concerns}
Our work showcases the feasibility and potential benefits of using generated images to facilitate online mental health self-disclosure. 
We have considered several potential ethical issues (e.g., privacy of participants) when choosing the style of generated images, developing MentalImager, and conducting user studies in a simulated community.
However, other ethical concerns should be considered for practical applications and follow-up research of our work. 
First, an intelligent tool like MentalImager should not force users to attach a generated image in their mental health self-disclosure posts if they do not feel like it. 
There are cases in which users have proper images on their local devices or do not want to attach images to their posts.
Users should have the control of turning off the image generation function. 

Second, as concerned by the participants (\autoref{subsubsec:perception}) in Study II the generated images may contain content or convey feelings that frighten users.
\zhenhui{There are tradeoffs between generating images based on the topical and emotional keywords of the mental health self-disclosure posts (what MentalImager does) and generating images with less negative content even though this content is disclosed in the posts. \citet{Li_Wu_Liu_Zhang_Guo_Peng_2024} informs the potential benefits of the formal case. 
It indicates that in online mental health communities, a post is also likely to get more social support if its text describes the visible content or tells a story depicted in the image or if the perceived emotions in the text and image are not in conflict. 
However, as people with mental health concerns often have weak psychological resilience, they could be susceptible to the negative content or emotion presented in the generated images. 
This could exacerbate the poster's mental health issues and isolate them from the online communities. 
As for the latter case, attaching an image with emotional conflict to the text or with a topic not directly matched with the text, is also common in online mental health communities, although it is not positively correlated to the received social support \cite{Li_Wu_Liu_Zhang_Guo_Peng_2024}. 
A safer version of MentalImager in the future could offer user-customized settings for guiding the model to generate images that have different relationships with the text in mental health self-disclosure posts. \citet{Li_Wu_Liu_Zhang_Guo_Peng_2024} has identified five such relationships regarding content (\ie visible, subjective, action, story, and meta) and three relationships in terms of emotion (\ie complement, dominance, conflict) in online mental health communities. We urge future work to implement and evaluate the models for generating images that have different relationships with the text for mental health self-disclosure.
Future applications of MentalImager should utilize necessary approaches to reduce the occurrences of generated harmful content and mitigate the negative impact if such content occurs. 
Besides, while we create a dictionary to exclude extremely negative keywords for image generation, we suggest future work to explore more methods to control and cope with the negativity of generated images. 
}
For example, the applications could build models to assess the degree of \zhenhui{negativity} of the generated images and provide useful links that help users cope with the feelings of being hurt by the generated content.

Third, people may misuse MentalImager to generate images that can invoke viewers' empathy but do not reveal the support-seekers' true experiences and feelings. 
To mitigate this issue, OMHCs could use AI techniques and human moderators to explicitly flag the potential AI-generated content in the support-seeking posts.
Such a flag could remind support-seekers to be honest in their posts and help support-providers to be aware of the possible deceptiveness of the generated images. 

\peng{
\subsection{Limitations and Future Work}

Our work has several limitations. 
First, we conducted our studies in short-term laboratory settings in a simulated OMHC rather than evaluating the long-term usage of MentalImager in a real OMHC. 
These choices could mitigate the uncontrollable impacts on people other than our participants in the process of exploring a new technology but could not eliminate the novelty effect. 
In the future, we plan to seek opinions on MentalImager from therapists and moderators of OMHCs to co-construct a safety prototype to test MentalImager in real OMHCs. 
Second, most of our participants \zhenhui{in Study II} are university students, which could not represent other specialized groups (such as pregnant women \cite{Gui17}). 
This confines our post topics to those pertinent to students \zhenhui{in Study II for the statistical comparison between an OMHC interface with MentalImager and a baseline interface without it}. 
\zhenhui{
Third, we measure the self-disclosure outcome from the support-providers' perspectives regarding their empathy with the support-seekers. It needs further explorations on how support-providers in a real online mental health community would perceive and reply to the posts created with the help of generated images.

Fourth, the detected one of the six emotional keywords in the emo-keyword-based prompt may not be able to reveal users' subtle emotions in their posts. 
While users can edit the keywords and adjust their weights to reveal any emotion to guide the image generator, future work could explore more methods to help users specify their subtle emotions. 
For example, future tools could provide an interactive arousal-valence map in which users can click any point to get a closely related emotional keyword for the image generator. 
Fifth, we assign the initial weights of keywords for the image generator based on trials and errors. While participants in Study II and III did not report the inappropriateness of these initial weights, a systematic study in the future could help better determine the initial weights of keywords. 
For example, the study could ask people to rate the topic- and emotion-relevance of the generated images given the same keywords with different weights. 
Sixth, while the keyword editor panel in MentalImager allows users to input keywords that match their characteristics and conditions, it could provide more options for users to specify these self-disclosure aspects. 
For example, it could provide drop-down selectors for users to select their gender and occupation as keywords for image generation. In case users face difficulties in manifesting their conditions, MentalImager could even leverage large language models to help users come up with related keywords. 
While the combination with LLMs to facilitate mental health self-disclosure has been out of the scope of this work, we highly recommend future work to explore this promising direction. 

Lastly, while we focus on facilitating text-based mental health self-disclosure in online communities via generated images, the generated images can also be used to talk about one's mental health outside the communities. 
For example, the generated images can be used to construct an emotional story of the user, which can facilitate people to reflect on their mental health conditions in offline scenarios, \eg in museum \cite{sensitive_pictures_chi22}. 
It would be also interesting to explore how the image generation model can facilitate online mental health self-disclosure via only images, which is also common in social media. 
For instance, without the need to draft a text-based post to get started, MentalImager could support users to directly manipulate the keywords to generate images and post them online. 
}

}

\section{Conclusion}
\peng{
To explore whether and how generative images can facilitate support-seekers in online mental health communities (OMHCs), we designed and presented MentalImager, a system that aids users in self-disclosure through generative images. 
Our within-subjects user study \zhenhui{II} with 24 participants shows that MentalImager can improve users' satisfaction with their self-disclosure in their support-seeking posts in a simulated online mental health community compared to a baseline interface without \zhenhui{MentalImager}. 
We also found that support-providers can better empathize with and are more willing to reply to the seekers via their posts created with MentalImager than those with the baseline interface. 
\zhenhui{
Another within-subjects study III with 30 participants further demonstrates that the generated images in MentalImager help users disclose themselves more clearly and reduce their worry about mental health self-disclosure. 
}
We report participants' concerns about the generated images for mental health self-disclosure and provide design considerations to facilitate support-seekers with generated images in OMHCs. 
Our work shows positive evidence for using generative techniques to enhance people's well-being and urges future work to improve these techniques for providing safer and more personalized support to people in need. 
}


\section{Appendix}
\zhenhui{
\subsection{List of Extremely Negative Keywords that Are Excluded for Image Generation}
sex, sexual, sexually,
nude, nudity, nudist
porn, pornography, pornographic, explicit, explicitly 
obscene, obscenity, erotic, erotica,
fetish, fetishes, fetishistic, fetishism, fetishist, fetishize,
masturbation, masturbate, masturbating, intercourse, intercourses,
orgasm, orgasms, orgasmic, penetration, penetrate, penetrating,
ejaculation, ejaculate, ejaculating,
incest, incestuous,
pedophile, pedophilia, pedophilic,
bestiality, bestial,
bondage, bondages,
domination, dominate, dominating,
submissive, submission,
sadism, sadist, sadistic,
masochism, masochist, masochistic,
BDSM, rape, raping, raped,
molestation, molest, molesting,
abuse, abusing, abused,
violence, violent, violently,
gore, gory,
torture, torturing, tortured,
murder, murdering, murdered, kill, killing, killed,
suicide, suicidal, self-harm, self-harming,
genital, genitals, penis, penises, vagina, vaginas,
clitoris, clitorises, anus, anuses, oral sex, oral, anal sex, anal,
rectum, rectal,
prostitute, prostitution, prostituting,
pimp, pimping,
brothel, brothels,
strip, stripping, stripper,
lap dance, lap dancing,
escort, escorting, fetishwear, fetishistic,
lingerie, lingeries,
orgy, orgies,
swinger, swinging, swinger party, swinger parties,
voyeur, voyeurism, voyeuristic,
exhibitionism, exhibitionist, exhibitionistic,
peep show, peep shows, camgirl, camboy, camming,
sexting, sexted,
dick pic, dick pics, nude photo, nude photos, sex tape, sex tapes,
gangbang, gangbanging, gangbanged,
cuckold, cuckolding, cuckolded,
MILF, MILFs, teen, teens, teenage,
barely legal,
cougar, cougars, sugar daddy, sugar daddies, sugar baby, sugar babies,
shemale, shemales, transsexual, transsexuals, crossdressing, crossdresser,
harem, harems
concubine, concubines, mistress, mistresses, infidelity, unfaithful, adultery, adulterous,
cheating, cheated,
rough sex, rough, 
fantasy, fantasies,
fuck, fucking, deviant, deviance,
hardcore, hard core, softcore, soft core,
skin flick, skin flicks,
peep show, peep shows,
sex toy, sex toys, dildo, dildos, vibrator, vibrators,
anal beads, butt plug, butt plugs,
cock ring, cock rings,bare sex doll,bare sex dolls, sex machine, sex machines,
glory hole, glory holes,
blood, bloody,
gore, gory,
violence, violent,
brutal, brutality,
weapon, weapons, fight, fighting,
assaulted, assaulting,assault,
murder, murderous,
attack, attacking, war, warfare, combat, battles, wound, wounded, injury, injuries, torture, torturing, pain, painful, suffering, suffer,
death, dead, die, dying,
slaughter, massacre,
execute, execution, executing,
decapitate, decapitation, behead, beheading,
hell, hellish,
damn, damned, damning, damnation,
curse, cursed, cursing,
damnable, damnably,
hellfire, inferno, infernal,
blasphemy, blasphemous, blaspheme,
doom, doomed, dooming,
wrath, wrathful, wrathfully,
hate, hated, hateful, hating,
rage, raging,
fury, furious, furiously,
violence, violent, violently,
brutal, brutality, brutalize, brutalized,
torment, tormented, tormenting,
torture, tortured, torturing
}

\end{document}